\documentclass[prx,reprint,a4paper,superscriptaddress,citeautoscript,floatfix]{revtex4-2}

\usepackage{amsmath,amsfonts,amssymb}
\usepackage{times}
\usepackage{mathrsfs}

\usepackage[pdftex]{graphicx}
\usepackage{microtype}
\usepackage{bm}\let\vec\bm

\usepackage[svgnames]{xcolor}
\usepackage[
	colorlinks=True,linkcolor=DarkRed,citecolor=ForestGreen,urlcolor=MediumBlue,
	pdfstartview=FitH,bookmarks=False,pdfpagemode=UseNone
]{hyperref}

\newcommand{\gqbar}{$\bar{g}_{\vec{q}}^2$}
\newcommand{\WS}{WS$_2$}
\newcommand{\WSe}{WSe$_2$}

\newif\ifshowcomments\showcommentstrue      
\begin{document}

\title{Electron-phonon coupling across the TMD/hBN van der Waals interface}

\author{G. Gatti} 
\affiliation{Department of Quantum Matter Physics, University of Geneva, 24 Quai Ernest-Ansermet, 1211 Geneva, Switzerland} 

\author{C. Berthod} 
\affiliation{Department of Quantum Matter Physics, University of Geneva, 24 Quai Ernest-Ansermet, 1211 Geneva, Switzerland}

\author{J. Issing} 
\affiliation{Department of Quantum Matter Physics, University of Geneva, 24 Quai Ernest-Ansermet, 1211 Geneva, Switzerland} 

\author{M. Straub} 
\affiliation{Department of Quantum Matter Physics, University of Geneva, 24 Quai Ernest-Ansermet, 1211 Geneva, Switzerland} 

\author{S. Mandloi} 
\affiliation{Department of Quantum Matter Physics, University of Geneva, 24 Quai Ernest-Ansermet, 1211 Geneva, Switzerland} 

\author{Y. Alexanian} 
\affiliation{Department of Quantum Matter Physics, University of Geneva, 24 Quai Ernest-Ansermet, 1211 Geneva, Switzerland} 

\author{J. Avila} 
\affiliation{Synchrotron SOLEIL, L’Orme des Merisiers, Saint Aubin-BP 48, 91192 Gif sur Yvette Cedex, France}

\author{P. Dudin} 
\affiliation{Synchrotron SOLEIL, L’Orme des Merisiers, Saint Aubin-BP 48, 91192 Gif sur Yvette Cedex, France}

\author{T. K. Kim} 
\affiliation{Diamond Light Source Ltd, Harwell Science and Innovation Campus, Didcot, OX11 0DE, United Kingdom}

\author{M. D. Watson} 
\affiliation{Diamond Light Source Ltd, Harwell Science and Innovation Campus, Didcot, OX11 0DE, United Kingdom}

\author{C. Cacho} 
\affiliation{Diamond Light Source Ltd, Harwell Science and Innovation Campus, Didcot, OX11 0DE, United Kingdom} 

\author{K. Watanabe}
\affiliation{Research Center for Electronic and Optical Materials, National Institute for Materials Science, 1-1 Namiki, Tsukuba, 305-0044, Japan}

\author{T. Taniguchi}
\affiliation{Research Center for Materials Nanoarchitectonics, National Institute for Materials Science, 1-1 Namiki, Tsukuba, 305-0044, Japan}

\author{W. Wang}
\affiliation{Department of Physics and Astronomy, University of Manchester, Manchester, United Kingdom}
\affiliation{National Graphene Institute, University of Manchester, Manchester, United Kingdom}
\affiliation{Henry Royce Institute for Advanced Materials, University of Manchester, Manchester, UK}

\author{N. Clark}
\affiliation{Department of Physics and Astronomy, University of Manchester, Manchester, United Kingdom}
\affiliation{National Graphene Institute, University of Manchester, Manchester, United Kingdom}
\affiliation{Henry Royce Institute for Advanced Materials, University of Manchester, Manchester, UK}

\author{R. Gorbachev}
\affiliation{Department of Physics and Astronomy, University of Manchester, Manchester, United Kingdom}
\affiliation{National Graphene Institute, University of Manchester, Manchester, United Kingdom}
\affiliation{Henry Royce Institute for Advanced Materials, University of Manchester, Manchester, UK}

\author{N. Ubrig} 
\affiliation{Department of Quantum Matter Physics, University of Geneva, 24 Quai Ernest-Ansermet, 1211 Geneva, Switzerland}
\affiliation{Group of Applied Physics, University of Geneva, 24 Quai Ernest Ansermet, 1211 Geneva, Switzerland}
\affiliation{POLIMA -- Center for Polariton-driven Light-Matter Interactions, University of Southern Denmark, Campusvej 55, DK-5230 Odense M, Denmark}

\author{I. Guti\'errez-Lezama} 
\affiliation{Department of Quantum Matter Physics, University of Geneva, 24 Quai Ernest-Ansermet, 1211 Geneva, Switzerland} 
\affiliation{Group of Applied Physics, University of Geneva, 24 Quai Ernest Ansermet, 1211 Geneva, Switzerland}

\author{A. Morpurgo} 
\affiliation{Department of Quantum Matter Physics, University of Geneva, 24 Quai Ernest-Ansermet, 1211 Geneva, Switzerland} 
\affiliation{Group of Applied Physics, University of Geneva, 24 Quai Ernest Ansermet, 1211 Geneva, Switzerland}
\author{A. Tamai} 
\affiliation{Department of Quantum Matter Physics, University of Geneva, 24 Quai Ernest-Ansermet, 1211 Geneva, Switzerland} 

\author{F. Baumberger}
\affiliation{Department of Quantum Matter Physics, University of Geneva, 24 Quai Ernest-Ansermet, 1211 Geneva, Switzerland} 
\affiliation{Swiss Light Source, Paul Scherrer Institut, 5232 Villigen PSI, Switzerland}

\date{\today}

\begin{abstract}
Many-body interactions can couple electronic states in one layer with collective excitations in the adjacent layer, providing a route to tailor properties of heterostructures. 
However, detecting and quantifying interlayer many-body interactions proved a major challenge.
Here, we demonstrate that quasiparticles in monolayer transition metal dichalcogenides (TMDs) are dressed by a remote cloud of phonons in the adjacent hexagonal boron nitride slab. Using angle resolved photoemission, we identify replica bands in the TMDs which are a clear fingerprint of long-range electron-phonon interaction. 
We develop a modified Fr\"ohlich model that shows semi-quantitative agreement with the experimental spectral functions.
Our analysis shows that remote electron-phonon coupling is a generic property of interfaces with hBN.
This has implications for electron mobilities in 2D materials, for superconductivity and possibly for moiré correlated phases.
\end{abstract}

\maketitle

Hexagonal boron nitride (hBN) is the substrate of choice for high-quality van der Waals heterostructures. Combining a large band gap with a chemically inert surface free of dangling bonds, hBN acts in many cases like a featureless dielectric that largely decouples the 2D material of interest from the environment~\cite{Dean2010,Yankowitz2019}.
Notable exceptions include moiré effects at interfaces of aligned hBN and graphene~\cite{Dean2013} or the static electric fields generated by artificial rhombohedral-stacking of hBN monolayers~\cite{Woods2021,Yasuda2021,Stern2021,Kiper2025}.
While these single-particle effects are well studied, much less is known about many-body interactions at interfaces with hBN.
Electron-phonon coupling (EPC) is of particular interest. EPC is responsible for a wide range of phenomena, many of which are important in 2D materials. It limits the ultimate carrier mobility~\cite{Ma2014,Sohier2021}, controls ultrafast carrier dynamics~\cite{Johannsen2013} and the cooling of hot carriers~\cite{Principi2017,Yang2018}, may be exploited to control metastable phases~\cite{Forst2011}, and can drive many-body instabilities including charge density waves~\cite{Zhu2015b} and superconductivity~\cite{Sohier2019}.

Interfacial many-body interactions first received broad attention when Allender, Bray and Bardeen proposed coupling of electrons in a metal film with excitonic fluctuations in the semiconductor substrate as a mechanism of superconductivity~\cite{Allender1973,Inkson1973}. 
Remote pairing at interfaces was also explored as a means to further enhance the critical temperature of cuprate superconductors~\cite{Kivelson2002,Gariglio2011}.
More recently, the high superconducting critical temperature of FeSe/SrTiO$_3$ was related to interfacial EPC detected in angle resolved photoemission (ARPES) experiments~\cite{Lee2014,Huang2017,Li2019}.

\begin{figure*}[t]
  \centering   \includegraphics[width=1\textwidth]{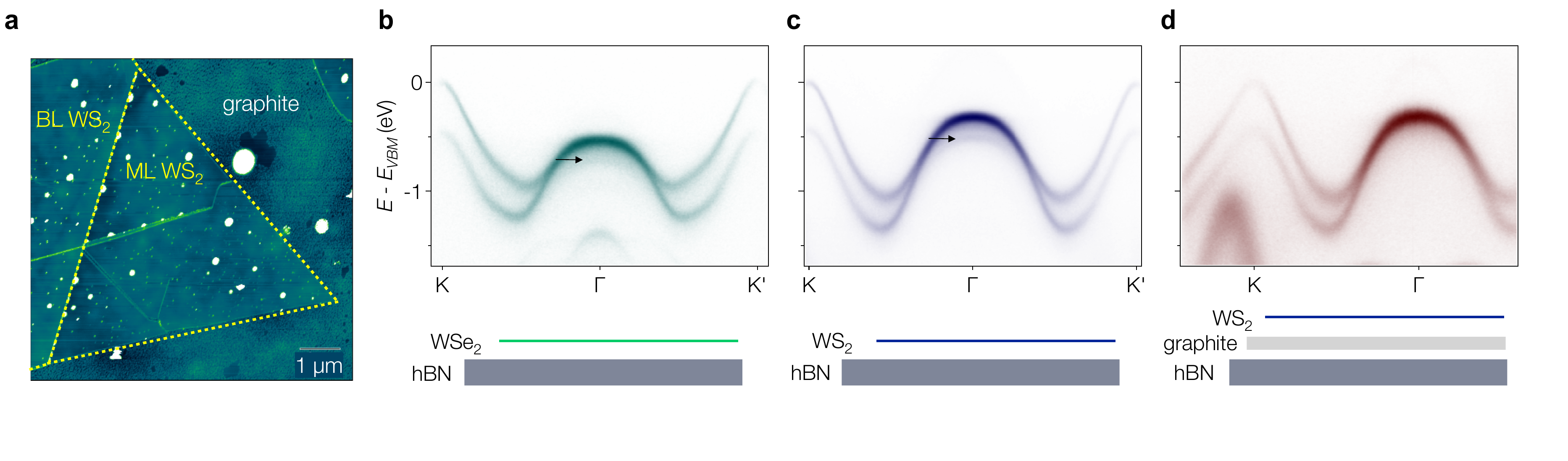}
  \caption[didascalia]{{\bf Replica bands from forward electron-phonon scattering.} 
  \textbf{a} Atomic force microscopy (AFM) topography map of a WS$_2$/graphite/hBN sample. \textbf{b}, \textbf{c} ARPES spectral functions along K-$\Gamma$-K' for monolayer WSe$_2$/hBN and WS$_2$/hBN. The black arrow marks the dominant replica band. \textbf{d} ARPES spectral function of WS$_2$/graphite/hBN. The replica band observed in (b,c) is absent.
 }
  \label{fig1}
\end{figure*}

Interfacial EPC may arise even with short-range coupling if lattice modes in one layer have an admixture of modes of the other layer. Such interfacial phonon hybridization was reported in heterostructures of covalent semiconductors and in oxide superlattices and is generally expected in the presence of strong chemical bonds if the phonon modes of both layers have similar frequencies~\cite{Chamberlain1993,Cheng2021,Driza2012}. These conditions, however, do not apply to the TMD/hBN interface.
Alternatively, EPC itself may be long range and couple quasiparticles in one layer with lattice modes of the adjacent layer. Long-range EPC is known from the Fr\"ohlich model, commonly employed to describe bulk polar insulators~\cite{Frohlich1954,Devreese2009}.
Long-range EPC across interfaces has been considered as a possible route to ultra-strong light matter interaction in cavities~\cite{Ashida2023} and was recently exploited for the all-electric excitation of hyperbolic phonon-polaritons~\cite{Guo2025}. It was further discussed as a potential upper limit for the mobility of metal-oxide semiconductor field-effect transistors (MOSFETs)~\cite{Chau2004} and of 2D materials on polar substrates~\cite{Fratini2008,Chen2008,Li2010,Ma2014,Vartanian2020,Sohier2021,Ponce2023}.
However, disentangling the numerous scattering processes that limit electron mobility remains a challenge~\cite{Ma2014} and direct spectroscopic evidence for long-range EPC across interfaces is scarce.
A recent photoemission study of graphene/hBN reported replicas of hBN valence band states, which were attributed to coupling across the interface to graphene phonons~\cite{Chen2018}. Yet, an unambiguous identification of an interface effect appears difficult considering that hBN itself is predicted to be polaronic and has phonons at the energies of the observed sidebands~\cite{Sio2023,Shahnazaryan2025}.
Raman scattering further reported signatures of hybridization of excitons in semiconducting transition metal dichalcogenides (TMDs) with phonons of hBN and other polar substrates~\cite{Jin2017,Chow2017,Merkl2021}.
These observations point to EPC across van der Waals interfaces.
However, a unifying picture of the mechanisms of interfacial EPC in 2D materials and of its effects on quasiparticle properties is lacking.

Here, we use micro-focus ARPES to map the spectral function of monolayer TMD semiconductors on hBN. Our data show replica bands in the TMDs at energies that far exceed their phonon frequencies but are comparable to lattice modes of the hBN substrate. This is a fingerprint of long-range interfacial electron-phonon coupling. We demonstrate semi-quantitative agreement with a generic model for electron propagation above a half space of hBN, suggesting that interfacial EPC is a universal property of interfaces of 2D semiconductors and hBN.

We study the prototypical systems of monolayer WS$_2$ and WSe$_2$ on hBN substrates with and without an intermediate graphite layer, as illustrated in Fig.~\ref{fig1}.
All samples were prepared from mechanically exfoliated flakes using dry transfer methods. hBN and graphite flakes were bulk-like with thickness $\sim 30$~nm and $\sim 15$~nm, respectively.
Figure~\ref{fig1}(a) shows a typical AFM topography map revealing a low density of bubbles and other defects, indicative of a good interfacial contact between layers. 
$\mu$-ARPES experiments have been performed at the I05 beamline of Diamond Light Source, and at the ANTARES beamline at Soleil synchrotron, at a fixed sample temperature of 180 K.
More details of the sample fabrication and ARPES experiments are given in Methods.

Figures~\ref{fig1}(b-d) show overview $\mu$-ARPES data for all three sample types. We observe clearly defined bands of the monolayer TMDs with a low background indicative of a good sample quality. 
The overall band structure with a single, spin degenerate band at the time-reversal invariant $\Gamma$ point and a pronounced spin-splitting at K is in agreement with band structure calculations~\cite{Zhu2011} and earlier ARPES studies~\cite{Wilson2017,Waldecker2019,Cucchi2021}.
However, a closer inspection of the data reveals additional spectral features that were not previously resolved. Below the $\Gamma$ valley, we find a satellite [marked by an arrow in panels (b, c)] that mimics the dispersion of the quasiparticle band near $\Gamma$ before merging with it at higher momenta. The dispersion plots further show hints of a second satellite in between the distinct satellite marked by an arrow and the main quasiparticle band.
Intriguingly, these satellites are present only for monolayer WS$_2$ and WSe$_2$ on hBN but not for WS$_2$ on graphite/hBN.

Replica bands in a spectral function are a fingerprint of electron-boson coupling strongly peaked at a wave vector $q=0$. In the simplest picture, such forward scattering causes sidebands at integer multiples of the boson energy $\hbar\omega_0$ with spectral weights following a characteristic Poisson distribution~\cite{Mishchenko2000}. 
Experimentally, signatures of forward scattering have been observed in low-density surface 2D electron liquids on polar oxides such as SrTiO$_3$, TiO$_2$ and EuO \cite{Wang2016,Moser2013,Riley2018}, in FeSe/SrTiO$_3$~\cite{Lee2014}, MoS$_2$/TiO$_2$~\cite{Xiang2023} and for electron-plasmon coupling in oxides and highly doped TMDs~\cite{Riley2018,Caruso2021,Jung2024,Ulstrup2024}.

\begin{figure*}[t!]
    \centering \includegraphics[width=1\textwidth]{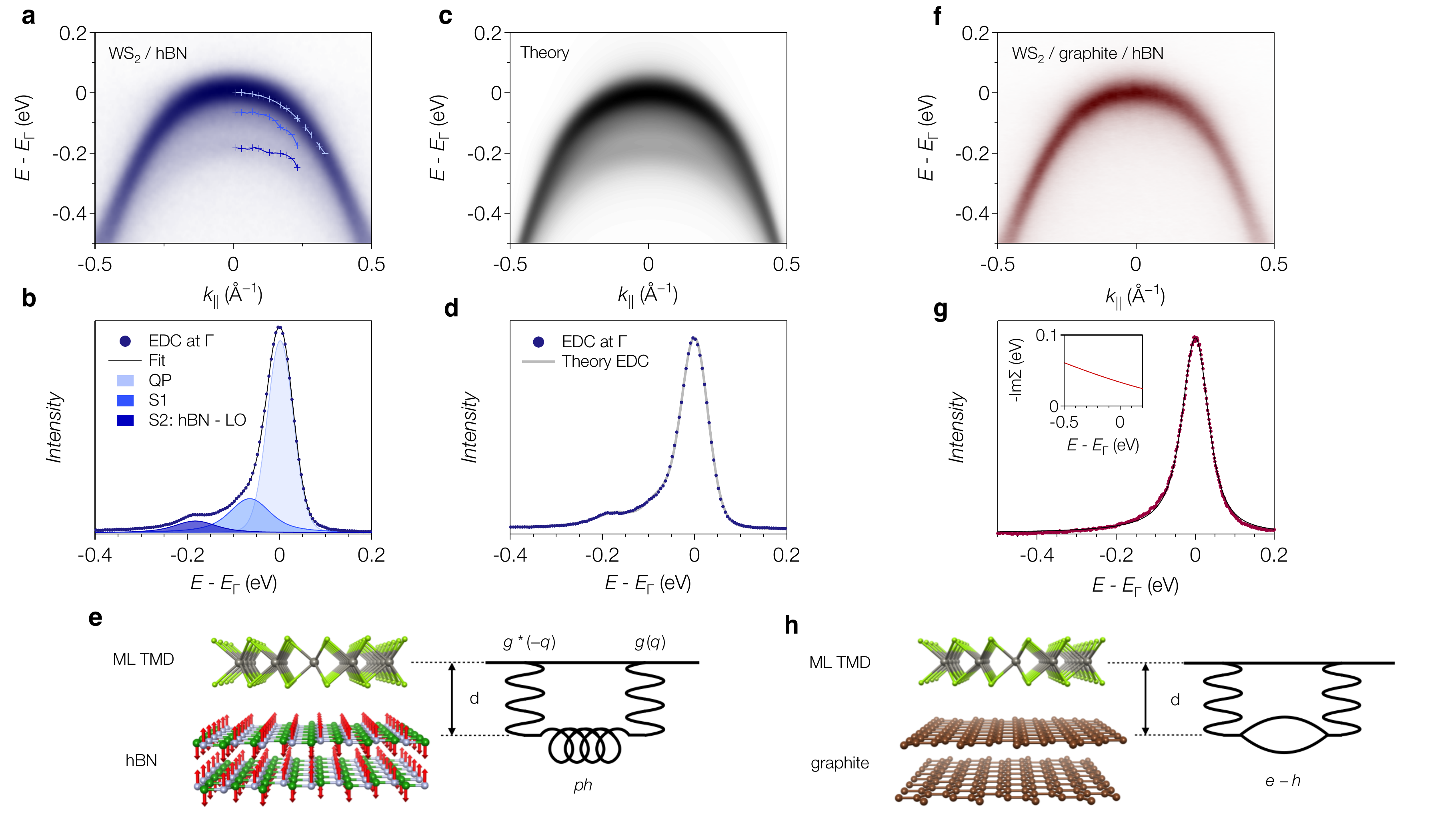}
  \caption{{\bf Spectral function analysis.}
  \textbf{a} $\Gamma$-valley ARPES spectral function of WS$_2$/hBN. Energies are given relative to the local VBM $E_{\Gamma}$ at $\Gamma$.
  Markers indicate the position of the QP, S1 and S2 components extracted from a fit with 3 components. 
  \textbf{b} EDC of WS$_2$/hBN taken at $\Gamma$ and a fit with the three components QP, S1 and S2. A small Shirley background has been subtracted from the data.
  \textbf{c} Fit with a minimal model for interfacial EPC. For details, see main text and App.~\ref{app:parameters}. 
  \textbf{d} Comparison of the model with the ARPES EDC at $\Gamma$.
  \textbf{e} Sketch of the WS$_2$/hBN interface and the associated Feynman diagram for remote EPC. 
  \textbf{f}, \textbf{g} $\Gamma$-valley ARPES spectral function of WS$_2$/graphite/hBN. The black line in \textbf{g} is a fit with the self-energy $\textrm{Im}\Sigma$ shown in the inset.  
  \textbf{h} Sketch of the WS$_2$/graphite interface with a diagram for coupling of quasiparticles to a polarization bubble in graphite.
    }
  \label{fig2}
\end{figure*}

In Fig.~\ref{fig2}, we focus on the energy / momentum region of the $\Gamma$ valley states of \WS/hBN. The energy distribution curve (EDC) at $\Gamma$ [Fig.~\ref{fig2}(b)] shows two shoulders, confirming the presence of two satellite bands. The EDC is well described by an empirical fit assuming a Lorentzian line shape convolved with a Gaussian for the main quasiparticle peak (QP) and the satellites S1, S2. This simple fit reveals key insight into the problem. First, we find the most pronounced satellite at $\sim 184$~meV far beyond the highest phonon frequencies of WS$_2$ of $\sim 55$~meV~\cite{Molina-Sanchez2011} but close to the in-plane optical modes of bulk hBN with energies of $170-200$~meV at $\Gamma$~\cite{Geick1966,Ponce2023,Sio2023}. Second, the fit gives spectral weights of $Z_{\textrm{QP}}\simeq 0.7$, $Z_{\textrm{S1}}\simeq 0.22$ and $Z_{\textrm{S2}}\simeq 0.08$ for the quasiparticle and the two satellites, respectively.
These weights deviate markedly from the Poisson distribution expected for multi-phonon processes, excluding an interpretation of S2 as a 2- or 3-phonon process.
Our data on \WSe/hBN shows the same key-features with a satellite at $\sim 180$~meV and a quasiparticle residue $\sim 0.7$.

The spectrum of WS$_2$/graphite/hBN shown in Fig.~\ref{fig2}(f, g) clearly deviates from WS$_2$/hBN.
Most prominently, satellite S2 is absent if WS$_2$ is placed on graphite. Instead, we find an asymmetric line shape with a long Lorentzian tail extending to high binding energy. This points to a decay of the WS$_2$ photohole via excitation of low-energy electron-hole pairs in the graphite substrate. Indeed, a fit with a generic Fermi liquid like self energy consistent with this decay channel (inset of Fig.~\ref{fig2}(g) and App.~\ref{app:FL}) provides an excellent description of the EDCs for WS$_2$/graphite/hBN. Together, these observations provide compelling evidence for coupling of WS$_2$ electrons to hBN phonons.

The energy of satellite S1 is less well defined. The simple 3-peak fit places it at $\sim 66$~meV, suggesting that it contains overlapping contributions from the optical phonons of WS$_2$ in the range of $43-55$~meV~\cite{Molina-Sanchez2011} and the out-of-plane optical (ZO) mode of hBN at 97~meV~\cite{Geick1966,Ponce2023,Sio2023}. We will thus focus the discussion primarily on the most distinct satellite S2.
We finally note that the line shapes in our TMD/hBN samples are close to Gaussian although they exceed the instrument resolution by nearly a factor of 2. This suggests that line widths are dominated by inhomogeneous broadening, likely arising from spatial variations of the chemical potential on a length scale below the $\mu$-ARPES beam spot of $\simeq 1$~$\mu$m.

We now outline a minimal model for the coupling of a generic 2D electron liquid (2DEL) with a polar substrate. Full details of the model are given in Apps.~\ref{app:self} and \ref{app:model}. We start from the Fr\"ohlich Hamiltonian 
\begin{equation}
H_{\mathrm{el-ph}}=\sum_{\vec{k}}\sum_{\vec{q}}g^{}_{\vec{q}}
	c^{\dagger}_{\vec{k}+\vec{q}_{\parallel}}c^{}_{\vec{k}}
	\left(b^{}_{\vec{q}}+b^{\dagger}_{-\vec{q}}\right),
\label{eq_main:H_int}
\end{equation}
where we restrict the electron wave vector $\vec{k}$ to 2D while the wave vector $\vec{q}$ and displacements of the phonons are 3D. Spin and phonon branch indices of the creation / annihilation operators of electrons $(c^{\dagger}/c^{})$ and phonons $(b^{\dagger}/b)$ have been omitted for simplicity. The matrix element $g_{\vec{q}}$ in Eq.~(\ref{eq_main:H_int}) derives from the interaction of the charge $e$ with the $1/r$ Coulomb potentials of the Born effective charges $Z$ of substrate lattice modes. 
This introduces a characteristic momentum dependence $e^{-qd}$ in the coupling function $g_{\vec{q}}$, where $d$ is the separation of the 2D electron system from the substrate surface. Integrating $g_{\vec{q}}$ over $q_z$ and assuming dispersionless phonons with a pure polarization, we obtain a 2D model with the new matrix element
\begin{equation}
\bar{g}_{\vec{q}}^2 \propto \alpha\,e^{-2qd}\,\frac{1-e^{-2qN_s c}}{1-e^{-2qc}},
 \label{eq_main:gbar2}
\end{equation}
where $\alpha$ is a dimensionless coupling constant, $N_s$ is the number of hBN layers and $c$ the interlayer distance in hBN.
For a semi-infinite substrate with $N_s\to\infty$, the coupling \gqbar{} behaves as $e^{-2qd}/qc$ for $q\to0$, leading to strong forward scattering with a maximum at $q=0$ and a decay length $1/d$.

Interfacial EPC is conceptually similar to the Fröhlich polaron model~\cite{Frohlich1950,Frohlich1954,Devreese2009,Mishchenko2000}. We highlight this similarity by defining the coupling constant $\alpha$ in Eq.~(\ref{eq_main:alpha}) such as to obtain a quasiparticle residue $Z=1-\alpha/2$, as it is obtained in the weak-coupling limit of the Fröhlich polaron model~\cite{Frohlich1950,Frohlich1954,Devreese2009,Mishchenko2000}. As shown in App.~\ref{app:Z}, this implies:
\begin{equation}
\alpha = \left(\frac{Z}{\epsilon/\epsilon_0}\right)^2\sqrt{\frac{m_b}{m}}
	\frac{C}{(\hbar\omega_0)^{5/2}}.
 \label{eq_main:alpha}
\end{equation}
Here, $\omega_0$ is the phonon frequency, $m_b$ and $m$ the band and electron masses, and $\epsilon/\epsilon_0$ the relative permittivity.
The constant $C$ depends only on the unit cell geometry and ionic masses of hBN, as detailed in App.~\ref{app:model}.

The quasiparticle spectral weight $Z_{\textrm{QP}}\simeq 0.7$ places our data in the weak-coupling regime, where the self-energy for electrons near a local valence band maximum with bare dispersion $\xi_{\vec{k}}$ is:
\begin{equation}
	\Sigma(\vec{k},E)=\int\frac{d^2q}{(2\pi)^2}\frac{\bar{g}_{\vec{q}}^2}
	{E-\xi_{\vec{k}+\vec{q}}+\hbar\omega_0+i\delta}.
    \label{eq_main:self}
\end{equation}
For a quantitative test of our minimal model of interfacial EPC, we focus on \WS/hBN and fit the experimental data with the spectral function
\begin{equation}
    A(\vec{k},E)=-\frac{1}{\pi}\mathrm{Im}\,
    \frac{1}{E-\xi_{\vec{k}}-\Sigma(\vec{k},E)}.
    \label{eq_main:Akw}
\end{equation}
Further analysis of the \WSe/hBN data is presented in App.~\ref{app:WSe2}.
Our model predicts coupling to the out-of-plane ZO mode of hBN with an energy of 97~meV and the LO mode at 200~meV in bulk and 170~meV in monolayer hBN, respectively~\cite{Sio2023,Sohier2021,Serrano2007}.
To correctly model satellite S1, we further include 2 optical phonons of WS$_2$ using a highly restricted model inspired by density functional perturbation theory results~\cite{Sohier2016}.
Full details of the analysis are given in Apps.~\ref{app:WS2}, \ref{app:parameters}, and \ref{app:k=0}.
The key parameters controlling the signatures of interfacial EPC are the hBN phonon frequencies with associated coupling constants and the structural parameters $N_s$, $d$ which control the momentum dependence of the matrix element \gqbar.
The latter are readily estimated from density functional theory (DFT) calculations and the thickness of hBN, to be $d\approx 5$~\AA{} and $N_s\approx 90$, respectively \cite{Magorrian2022}.
To test our model in an unbiased way, we nevertheless fit $N_s$ and $d$ to our ARPES data, such as to independently examine the predicted momentum dependence of \gqbar{} (see App.~\ref{app:parameters} for details of the fit).

The fit using this approach shown in Fig.~\ref{fig2}(c, d) is in excellent agreement with the data.
For interfacial EPC to the LO mode of hBN, the fit returns $\hbar\omega_{\textrm{LO}}=164$~meV, slightly lower than the simple 3-peak fit. This is consistent with diagrammatic Monte Carlo calculations, which find the onset of the satellite, rather than its maximum at the phonon frequency~\cite{Mishchenko2000}.
The coupling constant obtained from the fit is $\alpha_{\textrm{LO}}=0.17$, which is well in the weak-coupling regime, where the perturbative treatment used in Eq.~(\ref{eq_main:self}) is appropriate.
Importantly, the parameters defining $\alpha_{\textrm{LO}}$ can all be estimated independently, allowing for an independent cross-check. Using DFT values for the Born effective charge $Z_{\textrm{LO}}=2.71$ \cite{Ohba2001}, the dielectric constant $\epsilon_{\parallel} = 6.93$ \cite{Laturia-2018}, and the mass $m_b/m=2.83$ \cite{Ramasubramaniam-2012}, our model predicts $\alpha_2 = 0.24$, in good agreement with our analysis of the ARPES data.
For the ZO mode of hBN, our fit gives a coupling strength $\alpha_{\textrm{ZO}}=0.30$, very close to the value $0.29$ obtained from Eq.~(\ref{eq_main:alpha}) with $Z_{\textrm{ZO}}=0.82$ and $\epsilon_{\perp} = 3.76$~\cite{Ohba2001,Laturia-2018}.
Fits of the \WSe/hBN data are in good quantitative agreement with the above findings for \WS/hBN, confirming the universal nature of interfacial EPC (see App.~\ref{app:WSe2}).

\begin{figure*}[t!]
  \centering   \includegraphics[width=0.9\textwidth]{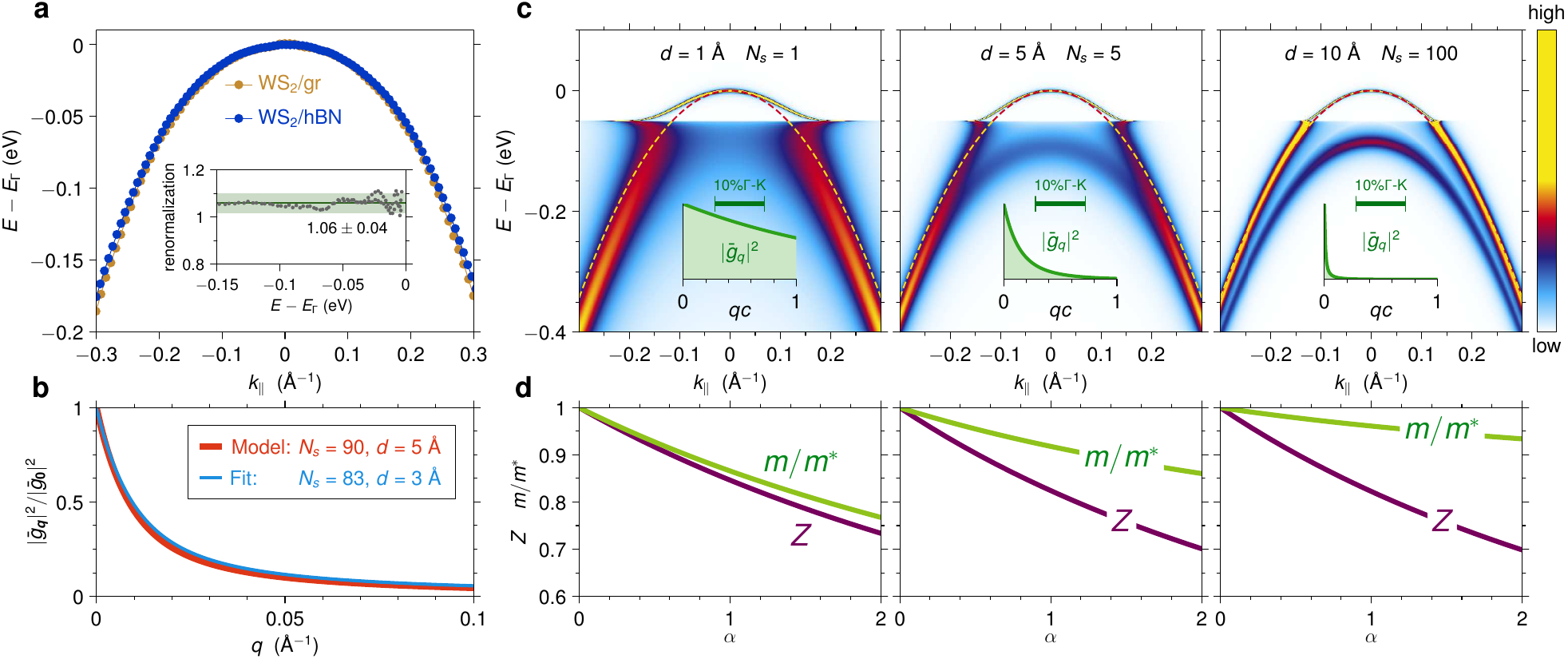}
  \caption[didascalia]{{\bf Suppression of mass enhancement and momentum dependence of interfacial EPC.}
  \textbf{a} Quasiparticle dispersion for WS$_2$/hBN and WS$_2$/graphite/hBN. The inset shows the renormalization $E-E_{\Gamma} \mathrm{(WS_2/gr)/}E-E_{\Gamma} \mathrm{(WS_2/hBN)}$, which is a measure of the mass enhancement $m^{*}/m$. \textbf{b} Comparison of \gqbar{} obtained from the best fit to the data with the prediction of Eq.~(\ref{eq_main:gbar2}) with the nominal parameters $N_s=90$, $d=5$~\AA.
  \textbf{c} Model spectral functions for different forms of \gqbar{} illustrate the evolution of kinks into replica bands with increasing forward scattering. All spectral functions are calculated for $\lambda=0.5$ and a single phonon mode with $\hbar\omega_0=50$~meV.
  \textbf{d} Mass enhancement $m^{*}/m$ and quasiparticle residue $Z$ for the model calculation in \textbf{c}.
 }
  \label{fig3}
\end{figure*}

EPC is commonly associated with mass enhancement. 
Indeed, for a self-energy that depends on energy only, typical for EPC in metals, one has $m^{*}/m=1+\lambda$, where $\lambda=-\partial\mathrm{Re}\,\Sigma/\partial E$.
Using the self-energies from our fit of the ARPES data, we find $\lambda_{\textrm{LO}}=0.048$ and $\lambda_{\textrm{ZO}}=0.091$ for interfacial coupling to the hBN modes.
Intriguingly though, our experiments indicate that mass enhancement is nearly absent for interfacial EPC. In Fig.~\ref{fig3}(a), we compare the low-energy quasiparticle dispersions of WS$_2$/hBN with WS$_2$/graphite/hBN, where interfacial EPC is quenched. From this we experimentally estimate $m^{*}/m=1.06\pm0.04$, below $1+\lambda_{\mathrm{hBN}}=1.14$.
This can be rationalized with the strong momentum dependence of interfacial EPC, which decouples the quasiparticle residue and effective mass.

Within our model, the forward nature of the coupling \gqbar{} is controlled by the distance $d$ between 2DEL and substrate and the number of substrate layers $N_s$ and increases with both parameters.
Model calculations for representative combinations of $N_s$ and $d$ shown in Fig.~\ref{fig3}(c) illustrate how the spectral function evolves from kinks in the dispersion, known from momentum independent coupling~\cite{Lanzara2001}, to the limit of clear replica bands for \gqbar{} very strongly peaked at $q=0$. At the same time, the low-energy mass enhancement, evident in the first panel with nearly uniform \gqbar, is suppressed with increasing forward scattering.
In App.~\ref{app:m*/m} we show that this suppression of mass enhancement in interfacial EPC arises from compensating effects of the energy and momentum dependence of the self-energy.

The qualitative changes of the spectral function with $N_s$ and $d$ further suggest that fitting these parameters provides a meaningful test of the predicted momentum dependence of interfacial EPC.
To this end, we compare in Fig.~\ref{fig3}(b) \gqbar{} from the best fit with the prediction obtained from the nominal values of the structural parameters $N_s$ and $d$. The excellent agreement confirms that our model correctly captures the nature of interfacial EPC.

Our data further constrain the nature and strength of EPC in monolayer WS$_2$.
The fit in Fig.~\ref{fig2}(c, d), includes two \WS{} phonon modes, one with momentum independent coupling and one with a generic forward interaction. This is inspired by density functional perturbation theory (DFPT) results that predict coupling with a weak momentum dependence to the out-of-plane $A_1$ mode of \WS{} with an energy of 51~meV and a forward interaction with the in-plane LO mode with an energy of 43~meV~\cite{Sohier2016}. In the fit shown in Fig.~\ref{fig2}(c, d), we fix the energy of both modes to the above values and use the matrix elements $g_{A_1}$ and $g_{\vec{q},\mathrm{LO}} =g_{\mathrm{LO}} \exp(-qr_0)$ with an inverse decay length $r_0=42$~\AA, which provides a good approximation of the DFPT results~\cite{Sohier2016}. With this approach, we obtain a coupling strength $g_{\mathrm{LO}}=1.7$~eV ($\lambda_{\mathrm{LO}}=0.23$) and negligible coupling to the $A_1$ mode.
We test the robustness of these values by varying the modes included in the fit as well as constraints on the mode energies (see App.~\ref{app:k=0}). This shows that a good description of the experimental data can only be obtained by including a forward-type interaction with a mode in the energy range of $40$--$60$~meV and a coupling in the range of $\lambda_{\mathrm{LO}}=0.15$--$0.2$. 
At the same time, we find that momentum-independent coupling to WS$_2$ optical phonons is small with an upper limit of $\lambda_{A_1}\approx 0.01$ to obtain a reasonable fit.
Notably, this behavior is opposite to DFPT results for isolated \WS{} monolayers, which show the strongest coupling for the $A_1$ mode. This calls for further theoretical work simulating realistic van der Waals interfaces rather than isolated monolayers.

Our work implies that forward scattering EPC is a universal property of interfaces with hBN or other polar insulators. 
This has implications ranging from quantum applications of hyperbolic phonon polaritons in hBN~\cite{Caldwell2014,Ashida2023} to key-properties of van der Waals heterostructures, including electron mobilities, superconductivity and moiré correlated phases~\cite{Fratini2008,Chen2008,Li2010,Ma2014,Vartanian2020,Sohier2021,Ponce2023,Jaoui2022,Wu2018b,Chen2024}.
It is interesting to note that the strength of interfacial EPC revealed here for TMD/hBN is comparable to FeSe/SrTiO$_3$, where $\lambda= 0.2$--$0.3$ was reported from ARPES experiments~\cite{Faeth2021, Wang2016_2}. The two systems further share the presence of a polar substrate. Moreover, superconductivity appears to be enhanced in both systems.
For gated WS$_2$/hBN, $T_c$'s up to 6~K were reported, $\sim 30\%$ above the highest $T_c$ on Si/SiO$_2$~\cite{Ding2022}. 
However, TMD/hBN interfaces also differ in important aspects from FeSe/SrTiO$_3$.
First, FeSe has strong covalent bonds with the SrTiO$_3$ substrate, suggesting that phonon hybridization plays a more important role than in van der Waals interfaces.
Second, the metallicity of FeSe screens long-range Coulomb interactions. Our model is thus expected to overestimate the coupling strength in FeSe/SrTiO$_3$. 
Indeed, assuming Fröhlich EPC inside SrTiO$_3$ with $\lambda\approx 3$, as reported in Ref.~\cite{Wang2016}, our model predicts interfacial EPC with $\lambda\approx0.6$, higher than the values deduced from ARPES experiments~\cite{Faeth2021, Wang2016_2}.

Our analysis further shows that the momentum dependence of EPC may be extracted quantitatively from ARPES data. We anticipate rapid progress in this regard from improved data quality and new analysis methods. This promises a wealth of new insight into problems as diverse as the quantum efficiency of LEDs~\cite{Jhalani2017,Waldecker2017}, the stability of charge density waves~\cite{Johannes2008,Zhu2015b}, or unconventional superconductivity in systems ranging from cuprates~\cite{Bulut1996,Johnston2010} to atomically thin layers such as FeSe/SrTiO$_3$~\cite{Rademaker2016,Lee2014} and correlated moiré superlattices~\cite{Wu2018b,Chen2024}.

\begin{acknowledgments}
We acknowledge discussions with M. Berciu, O. Barišić, E. Demler, K. Ensslin, M. Gibertini, A. Imamoglu, A.S. Mishchenko, S. Ponc\'e, M. LeTacon.
This work was supported by the Swiss National Science Foundation (SNSF) under grants 184998, 178891.
RG acknowledges support of ERC Consolidator grant QTWIST (no. 101001515), European Quantum Flagship Project 2DSIPC (no. 820378) and EPSRC (grant numbers EP/V007033/1, EP/V026496/1, EP/S030719/1  and EP/Z531121/1).
KW and TT acknowledge support from the JSPS KAKENHI (Grant Numbers 21H05233 and 23H02052), the CREST (JPMJCR24A5), JST and World Premier International Research Center Initiative (WPI), MEXT, Japan.
We acknowledge Diamond Light Source for time on Beamline I05-1 under Proposals SI29021, SI31915, SI36214. 
We acknowledge Soleil Synchrotron for time on Beamline ANTARES under Proposal 20240524.
\end{acknowledgments}

\appendix

\section{Experimental methods}
\label{app:methods}

Single crystals of hBN, WS$_2$ and WSe$_2$ have been exfoliated in air on 285~nm SiO$_2$/Si wafers. Crystal flakes were identified from their optical contrast under a microscope, and their thickness was later confirmed by AFM measurements. Heterostructures were assembled in a glove box using a motorized transfer system.
The TMD/hBN heterostructures used to obtain the data shown in this work 
were fabricated using polymer transfer stamps, while the WS$_2$/graphite/hBN heterostructure was assembled using a metal-coated SiN cantilever.
Polymer stamps were fabricated by placing a spin-coated polypropylene carbonate (PPC) film on a dome-shaped handle made of polydimethylsiloxane (PDMS). 
The different flakes of the TMD/hBN heterostructures were stacked in inverted order (namely starting from hBN), then flipped and released on a graphetized SiC substrate.
Polymer residues and other contaminants were removed by annealing in high vacuum at 350$^{\circ}$C for 2.5 hours.
Metal-coated SiN cantilever were prepared at Manchester University with the method described in Ref.~\cite{Wang:2022aa,Wang2023} and were shipped under protective atmosphere to the University of Geneva. Prior to use, they were etched lightly in an argon - oxygen plasma. The flakes of the WS$_2$/graphite/hBN heterostructure were again stacked in inverted order starting from hBN. The SiN cantilever was subsequently broken off its substrate and glued on the graphetized surface of a SiC wafer using conductive silver epoxy.

Photoemission experiments were performed at the nanoARPES endstation of the I05 beamline at Diamond Light Source, Didcot (UK) and the Antares beamline of Soleil synchrotron, France. 
The synchrotron beam was focused to $\approx 1\times1$~$\mu$m$^2$ spot size using a Fresnel zone plate. 
The WS$_2$/hBN and WSe$_2$/hBN samples were measured at a photon energy of 80~eV, while the WS$_2$/graphite/hBN sample was measured using 95~eV light. All experiments used $p$-polarized light. 
Immediately prior to the experiments, the sample was annealed in ultra-high vacuum at 400$^\circ$C for 90 minutes before transferring it \textit{in-situ} to the experimental chamber. The photoemission experiments were performed at 180~K using an energy and momentum resolution of $20-30$~meV and $\approx 0.01$~\AA$^{-1}$. 
Note that the WSe$_2$ and WS$_2$ flakes were not in direct electrical contact with the chamber ground. Nevertheless, charging effects remained well below the resolution of this experiment, which we attribute to the high photo-conductivity of hBN.

\section{Interaction of two-dimensional electrons with a polar substrate}
\label{app:self}

Consider a system hosting two-dimensional electrons---thereafter abbreviated as 2DEG for convenience---lying at a distance $d$ above a polar substrate. The substrate is modeled as a crystal of ions, each characterized by an effective charge $Z_{\nu}$ and an ionic mass $M_{\nu}$. The ions are located at positions $\vec{R}+\vec{\tau}_{\nu}+\vec{u}_{\nu}$, where $\vec{R}$ represents any Bravais lattice vector in the half-space $z\leqslant 0$, $\vec{\tau}_{\nu}$ gives the equilibrium position of the ion in the unit cell, and $\vec{u}_{\nu}$ is the displacement corresponding to the phonons of the crystal ($\vec{u}_{\nu}$ depends on $\vec{R}$). Each ion carries a long-range potential
\begin{equation}
	V_{\nu}(\vec{x})=\frac{eZ_{\nu}}{4\pi\epsilon|\vec{x}|},
\end{equation}
where $\epsilon$ represents the screening of the electric field generated by the ionic motion. The potential felt by the electrons in the 2DEG is $V(\vec{r})=\sum_{\vec{R}\nu}V_{\nu}(\vec{r}-\vec{R}-\vec{\tau}_{\nu}-\vec{u}_{\nu})$. We assume that the effect of the equilibrium substrate potential, obtained by setting $\vec{u}_{\nu}=\vec{0}$, is already incorporated in the band structure of the 2DEG. The interaction of the 2DEG electrons with the substrate phonons is given by the change of $V(\vec{r})$ at first-order in the displacements, which couples to the electronic charge density $-en(\vec{r})$ according to
\begin{equation}
	H_{\mathrm{el-ph}}=e\int d^2r\,n(\vec{r})
	\sum_{\vec{R}\nu}\vec{u}_{\nu}\cdot\vec{\nabla}V_{\nu}(\vec{r}-\vec{R}-\vec{\tau}_{\nu}).
\end{equation}
We follow a standard procedure \cite{Mahan-2000, Bruus-2016, Berthod-2018} to rewrite this Hamiltonian in second-quantized form. The density $n(\vec{r})$ is expressed in terms of the electron creation/annihilation operators $c^{\dagger}_{\vec{k}\sigma}$/$c^{}_{\vec{k}\sigma}$ and the displacements $\vec{u}_{\nu}$ in terms of the phonon operators $b^{}_{\vec{p}s}+b^{\dagger}_{-\vec{p}s}$, where $\sigma$ labels the electronic spins and $s$ the various phonon branches with frequencies $\omega_{\vec{p}s}$. The only difference with respect to the textbook descriptions is that, here, the position vector $\vec{r}$ and the corresponding wave-vector $\vec{k}$ are two-dimensional (2D) vectors belonging to the 2DEG, while $\vec{R}$, $\vec{\tau}_{\nu}$, $\vec{u}_{\nu}$, and the wave-vector $\vec{p}$ are three-dimensional (3D) vectors of the polar substrate. After introducing the in-plane 2D Fourier transform of the ionic potential,
\begin{equation}
	V_{\nu}(\vec{k},z)=\int d^2r\,V_{\nu}(\vec{r},z)e^{-i\vec{k}\cdot\vec{r}}
	=\frac{eZ_{\nu}}{2\epsilon}\frac{e^{-k|z|}}{k},
\end{equation}
we arrive at the usual form of electron-phonon Hamiltonian,
\begin{equation}\label{eq:Hel-ph}
	H_{\mathrm{el-ph}}=\sum_{\vec{k}\sigma}\sum_{\vec{p}s}g^{}_{\vec{p}s}
	c^{\dagger}_{\vec{k}+\vec{p}_{\parallel}\sigma}c^{}_{\vec{k}\sigma}
	\left(b^{}_{\vec{p}s}+b^{\dagger}_{-\vec{p}s}\right).
\end{equation}
The electron scattering is from $\vec{k}$ to $\vec{k}+\vec{p}_{\parallel}$, because the 3D phonons can only exchange \emph{in-plane} momentum with the 2DEG electrons. Note that we used a plane-wave basis for both the electron and the phonon operators, and as a result the matrix element is independent of the electron wave-vector $\vec{k}$:
\begin{multline}\label{eq:g}
	g_{\vec{p}s}=\frac{ie^2}{2\epsilon S_{\mathrm{cell}}}
	\sum_{\nu}Z_{\nu}\sqrt{\frac{\hbar}{2M_{\nu}\omega_{\vec{p}s}N}}
	e^{-i\vec{p}_{\parallel}\cdot\vec{\tau}_{\nu}}\\
	\times\left(\vec{e}_{\vec{p}s}^{\nu}\cdot\hat{\vec{p}}_{\parallel}
	+i\vec{e}_{\vec{p}s}^{\nu}\cdot\hat{\vec{z}}\right)
	\sum_{n=0}^{N_s-1}e^{ip_zz_n}e^{-p_{\parallel}|d-z_n-\tau_{\nu}^z|}.
\end{multline}
$S_{\mathrm{cell}}$ is the in-plane surface of the substrate unit cell, $N$ the total number of unit cells in the substrate, $\vec{e}_{\vec{p}s}^{\nu}$ is the dimensionless 3D displacement of the ion $\nu$ for the phonon with quantum numbers $\vec{p}$ and $s$, $\hat{\vec{p}}_{\parallel}=\vec{p}_{\parallel}/p_{\parallel}$, $\hat{\vec{z}}$ is the unit vector in the $z$ direction, $N_s$ is the number of layers in the substrate, and $z_n=-nc$ with $c$ the lattice parameter of the substrate in the $z$ direction. This expression shows that the 2DEG electrons couple to all substrate phonons, except those that are both in-plane ($\vec{e}_{\vec{p}s}^{\nu}\perp\hat{\vec{z}}$) and purely transverse ($\vec{e}_{\vec{p}s}^{\nu}\perp\vec{p}$).

At leading order, the Hamiltonian (\ref{eq:Hel-ph}) renormalizes the single-electron retarded Green's function in the 2DEG with the self-energy \cite{Mahan-2000, Coleman-2015, Bruus-2016, Berthod-2018}
\begin{equation}\label{eq:self-energy}
	\Sigma(\vec{k},E)=\sum_{\vec{p}s\pm}|g_{\vec{p}s}|^2\frac{f(\pm\xi_{\vec{k}+\vec{p}_{\parallel}})
	+b(\hbar\omega_{\vec{p}s})}{E-\xi_{\vec{k}+\vec{p}_{\parallel}}\pm\hbar\omega_{\vec{p}s}+i\delta}.
\end{equation}
$f$ and $b$ are the Fermi-Dirac and Bose-Einstein distributions, respectively, $\xi_{\vec{k}}$ is the 2DEG dispersion relation measured from the chemical potential, and $\delta$ is a positive infinitesimal. A complete evaluation of Eq.~(\ref{eq:self-energy}) requires knowing the 2DEG dispersion relation and the phonons of the model 3D ionic substrate, which is outside the scope of the present work. Below, we specialize Eq.~(\ref{eq:self-energy}) to the case of a single dispersionless phonon mode. We furthermore take into account the particular situation of WS$_2$ and WSe$_2$ near the $\Gamma$ point, where the 2DEG dispersion is a fully occupied hole band deeply below the chemical potential.

\section{\boldmath Self-energy model for a semi-conducting TMD near the $\Gamma$ point and a dispersionless hBN phonon}
\label{app:model}

In the case of WS$_2$ and WSe$_2$ monolayers deposited on hBN, the hole band near the $\Gamma$ point lies $\sim 0.5$~eV below the chemical potential. Considering only this region of momentum space, we can set $f(\xi_{\vec{k}+\vec{p}_{\parallel}})=1$ and $f(-\xi_{\vec{k}+\vec{p}_{\parallel}})=0$. The Bose-Einstein factor may also be dropped for the phonons of interest with energies $\gtrsim50$~meV, much larger than the thermal energy ($\sim 15$~meV). Equation~(\ref{eq:g}) simplifies drastically for a dispersionless phonon, as the $\vec{p}$ dependencies of $\omega_{\vec{p}s}$ and $\vec{e}_{\vec{p}s}^{\nu}$, disappear. For the semiconducting monolayers of interest and a single dispersionless substrate phonon of frequency $\omega_0$, the model therefore becomes
\begin{equation}\label{eq:Sigma}
	\Sigma(\vec{k},E)=\int\frac{d^2q}{(2\pi)^2}\frac{\bar{g}_{\vec{q}}^2}
	{E-\xi_{\vec{k}+\vec{q}}+\hbar\omega_0+i\delta},
\end{equation}
where $\vec{q}$ is a 2D vector and the new matrix element absorbs the sum over $p_z$, $\bar{g}_{\vec{q}}^2=N_{\parallel}S_{\mathrm{cell}}\sum_{p_z}|g_{\vec{q},p_z}|^2$. If we set the origin of coordinates between the Boron and Nitrogen atoms, the atomic positions are $\vec{\tau}_1=-\vec{\tau}_2=\vec{\tau}/2$, with $|\vec{\tau}|=1.44$~\AA{} the equilibrium Boron--Nitrogen distance, and since both atoms are in plane we have $\tau_{\nu}^z=0$. We consider for definiteness a dipolar motion of the two atoms in the direction $\vec{n}$, by setting $Z_1=-Z_2\equiv Z$ and $\vec{e}_{\vec{p}}^{\nu=1}=-\vec{e}_{\vec{p}}^{\nu=2}\equiv\vec{n}/\sqrt{2}$. The resulting expression of the matrix element is
\begin{multline}\label{eq:gbar}
	\bar{g}_{\vec{q}}^2=\left(\frac{Ze^2}{4\epsilon}\right)^2\frac{1}{S_{\mathrm{cell}}}
	\left(\frac{1}{\sqrt{M_1}}+\frac{1}{\sqrt{M_2}}\right)^2\frac{\hbar}{\omega_0}\\
	\times|\vec{n}\cdot\hat{\vec{q}}+in_z|^2e^{-2qd}\frac{1-e^{-2qN_sc}}{1-e^{-2qc}}.
\end{multline}
The approximation $\cos(\vec{q}\cdot\vec{\tau})\approx 1$ has been used, which is justified in the region $q<1/|\vec{\tau}|$ of interest, where the $q$ dependence is dominated by the exponential factors in Eq.~(\ref{eq:gbar}).

The two relevant zone-center phonons of hBN are the ZO phonon with frequency $\sim 23~\mathrm{THz}$, which corresponds to out-of-plane motion, and the LO phonon with frequency $\sim 41~\mathrm{THz}$ corresponding to in-plane motion. A mode that is purely out-of-plane has $\vec{n}\cdot\hat{\vec{q}}=0$ and $n_z=1$, while an in-plane mode that is purely longitudinal has $\vec{n}\cdot\hat{\vec{q}}=1$ and $n_z=0$. Assuming an ideal polarization, we can therefore set $|\vec{n}\cdot\hat{\vec{q}}+in_z|^2=1$ for both the ZO and LO modes of hBN. For convenience, we write the electron-phonon matrix element as
\begin{equation}\label{eq:gbar1}
	\bar{g}_{\vec{q}}^2=\alpha\sqrt{\frac{32\hbar^5c^2\omega_0^3}{m_b}}
	e^{-2qd}\frac{1-e^{-2qN_sc}}{1-e^{-2qc}},
\end{equation}
where the parameter $\alpha$ is, using the known values for hBN,
\begin{align}\label{eq:alphabar}
	\nonumber
	\alpha&=\left(\frac{Ze^2}{4\epsilon}\right)^2\frac{1}{S_{\mathrm{cell}}}
	\left(\frac{1}{\sqrt{M_1}}+\frac{1}{\sqrt{M_2}}\right)^2\sqrt{\frac{m_b}{32\hbar^3c^2\omega_0^5}}\\
	&=\left(\frac{Z}{\epsilon/\epsilon_0}\right)^2\sqrt{\frac{m_b}{m}}
	\frac{0.01~\mathrm{eV}^{5/2}}{(\hbar\omega_0)^{5/2}}.
\end{align}
With the normalization chosen in Eq.~(\ref{eq:gbar1}), where $m_b$ is the mass of the hole band and $m$ the electron mass, the parameter $\alpha$ is dimensionless and plays the same role as the dimensionless coupling of the Fr\"{o}hlich polaron model \cite{Mishchenko2000} (see App.~\ref{app:Z}). The peculiar momentum dependence in Eq.~(\ref{eq:gbar1}) is the main feature of this model. For a bulk substrate with $N_s=\infty$, the coupling behaves as $e^{-2qd}/(2qc)$ for $q\to0$, which leads to a strong forward scattering effect with maximum scattering at $q=0$. In the monolayer limit $N_s=1$, the $1/q$ enhancement is lost and the coupling behaves as $e^{-2qd}$, leading to a suppression of the coupling at the zone center and a maximum at $qd=1/2$, because of the factor $q$ coming from the radial integral in Eq.~(\ref{eq:Sigma}).

When comparing the model with experimental data, we use a hole band with a quartic correction that we write as $\xi_{\vec{k}}=E_{\Gamma}-\frac{\hbar^2k^2}{2m_b}(1+Sk^2)$. Furthermore, in our simulations we remove the constant value $\mathrm{Re}\,\Sigma(\vec{0},E_{\Gamma})$ from the self-energy Eq.~(\ref{eq:Sigma}), i.e., we assume that the real part of the self-energy vanishes at the top of the valence band, whose energy is, therefore, not renormalized. The Born effective charges of hBN were calculated in Ref.~\cite{Ohba-2001} and found to be different for displacements parallel and perpendicular to the $c$ axis. We use the reported values for the out- and in-plane phonons, e.g., $Z_{\mathrm{ZO}}=0.82$ and $Z_{\mathrm{LO}}=2.71$. The dielectric screening is also different for displacements parallel and perpendicular to the $c$ axis, and should be evaluated at the frequency of the phonon. Here, we consider the phonon frequencies $\omega_{\mathrm{ZO}}$ and $\omega_{\mathrm{LO}}$ as well as $\epsilon_{\mathrm{ZO}}$ and $\epsilon_{\mathrm{LO}}$ as adjustable parameters.

\section{\boldmath Self-energy model for the coupling to WS$_2$ phonons}
\label{app:WS2}

In the free-standing WS$_2$ monolayer, the electrons near the $\Gamma$ point interact mostly with a LO phonon of energy $43$~meV and an $A_1$ phonon of energy $51$~meV. The coupling to the latter is approximately momentum independent, while the coupling to the former is strongly enhanced near $q=0$ \cite{Sohier2016}. The minimal self-energy model describing these relaxation processes has the form of Eq.~(\ref{eq:Sigma}), however with $\bar{g}_{\vec{q}}$ replaced by a different coupling function. We use a constant coupling $g_{A_1}$ for the $A_1$ mode and a coupling $g_{\vec{q},\mathrm{LO}}(q)=g_{\mathrm{LO}}e^{-qr_0}$ with $r_0=42$~\AA{} for the LO mode, consistently with the first-principles calculations \cite{Sohier2016}. We treat $g_{A_1}$ and $g_{\mathrm{LO}}$ as adjustable parameters, but we freeze the phonon frequencies to the values $43$ and $51$~meV. The first-principles values of $g_{A_1}$ and $g_{\mathrm{LO}}$ are close to $0.09$~eV and $0.165$~eV, respectively.

\section{\boldmath Estimation and optimization of the model parameters for WS$_2$ on hBN}
\label{app:parameters}

The initial values of the model parameters are estimated as follows. The maximum of the EDC at $k_{\parallel}=0$ provides $E_{\Gamma}$, the top of the valence band, that we set hereafter to zero, $E_{\Gamma}$ serving as the origin of the energy axis. We determine the band mass $m_b\approx2.33m$ and the quartic correction $S\approx3.09$~\AA{}$^2$ from a fit to the theoretical band calculated within the GW approximation in Ref.~\cite{Ramasubramaniam-2012}. The distance to the hBN substrate is estimated as $d\approx5$~\AA{}, which corresponds to the calculated distance between the hBN top layer and the $\Gamma$-point states located in the tungsten layer \cite{Magorrian2022}, while the number $N_s\approx90$ of hBN layers corresponds to the estimated substrate thickness of $\sim 30$~nm. The energies of the two hBN phonons are set to $\hbar\omega_1\approx97$~meV and $\hbar\omega_2\approx170$~meV, as reported in Ref.~\cite{Serrano-2007} for the ZO and LO/TO modes of bulk hBN at the $\Gamma$ point, respectively. The relative permittivity $\epsilon$ is more difficult to estimate, as it should in principle represent the cooperative screening by the hBN substrate and WS$_2$ monolayer at the frequencies of the phonons. To begin with an order-of-magnitude estimate, we ignore the screening of WS$_2$ and we use the bulk hBN static permittivities calculated theoretically in Ref.~\cite{Laturia-2018}. The reported values are 3.76 for an out-of-plane field and 6.93 for an in-plane field. We therefore set $\epsilon_1\approx3.8$ for the out-of-plane hBN ZO mode and $\epsilon_2\approx6.9$ for the in-plane hBN LO/TO mode. For the two WS$_2$ phonons, we use the first-principles values given in the previous section. The last two model parameters are a small residual scattering rate tentatively fixed to $\Gamma\approx5$~meV and an experimental Gaussian broadening, determined in Fig.~\ref{fig2}(b) of the main text to be $\Delta E\approx64$~meV.

\begin{figure}[tb]
\includegraphics[width=\columnwidth]{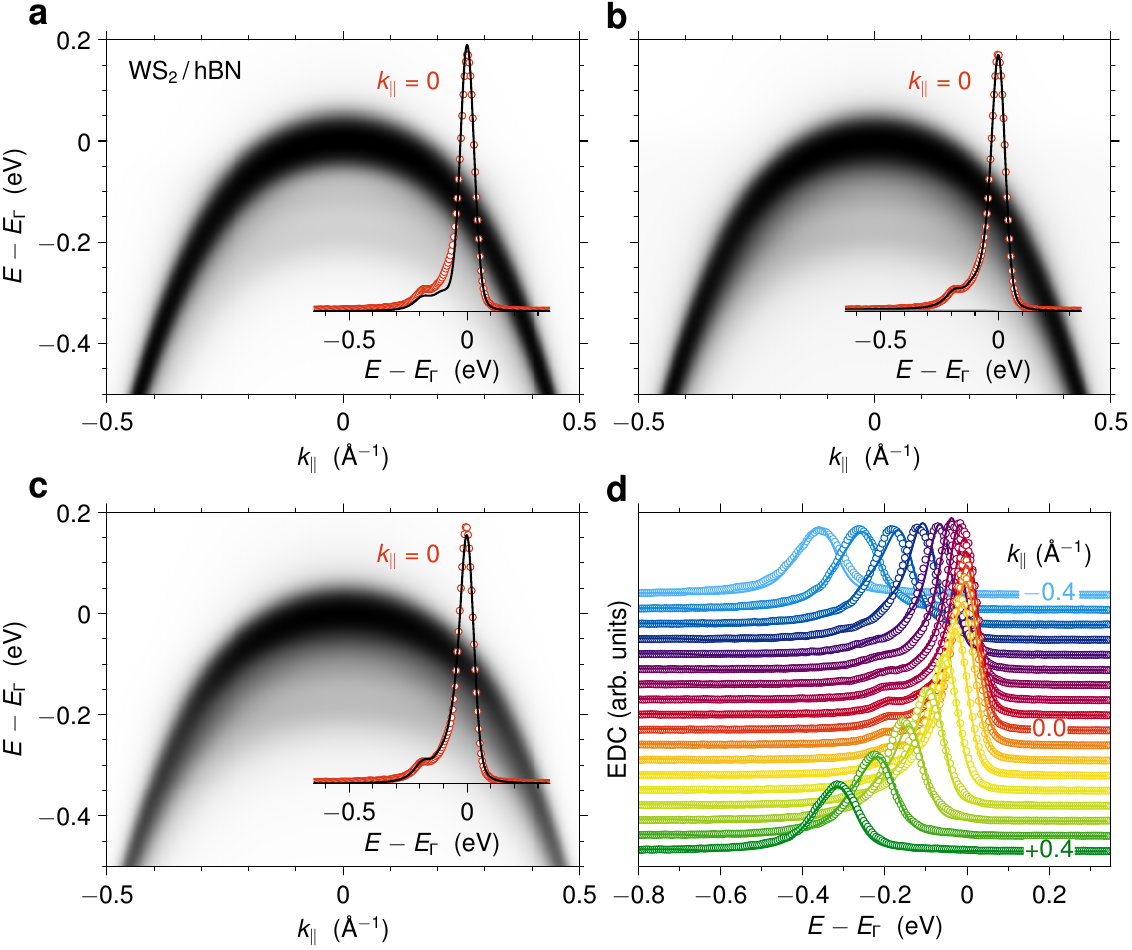}
\caption{\label{fig:parameters}
\textbf{a} Theoretical ARPES bandmap computed with the initial values of the parameters. The inset shows the experimental EDC at $k_{\parallel}=0$ (red symbols) together with the theoretical curve. \textbf{b} ARPES bandmap computed after some parameters have been adjusted to fit the EDC at $k_{\parallel}=0$. A small background (grey shade) is included. \textbf{c} ARPES bandmap and EDC at $k_{\parallel}=0$ computed after an optimization of all parameters. \textbf{d} Best fit (solid lines) to 18 experimental EDC curves (circles) with a momentum dependent amplitude. The curves are shifted vertically for clarity.
}
\end{figure}

The prediction of the model based on these initial parameters for coupling to hBN and WS$_2$ phonons is displayed in Fig.~\ref{fig:parameters}(a). The two satellites are clearly resolved, showing that the model provides the correct order of magnitude for the overall coupling strength, without a need to fine tune the parameters. For a first quick optimization, we select the EDC at $k_{\parallel}=0$ and fit the parameters $d$, $\hbar\omega_2$, $\epsilon_{1,2}$, $g_{\mathrm{A_1}}$, $g_{\mathrm{LO}}$, $\Gamma$, and $\Delta E$, while keeping the ratio $\hbar\omega_1/\hbar\omega_2$ fixed to the value $97/170$. At this point, we find that the best fit largely removes the isotropic coupling to the $A_1$ WS$_2$ phonon and boosts the coupling to the WS$_2$ LO phonon. The resulting parameters are $d=4.5$~\AA{}, $\hbar\omega_2=168$~meV, $\epsilon_{1,2}=3.2, 6.8$, $g_{\mathrm{A_1}}=0.01$~eV, $g_{\mathrm{LO}}=1.5$~eV, $\Gamma=7$~meV, and $\Delta E=59$~meV. For this fit, we also introduced a background of the form $B_1/\{\exp[(E-B_2)/B_3]+1\}$, which is found to provide a marginal contribution of relative amplitude $B_1=0.08~\mathrm{eV}^{-1}$, with $B_2=-67$~meV and $B_3=17$~meV [shaded gray curve in the inset of Fig.~\ref{fig:parameters}(b)]. With these adjusted parameters, the predicted line-shape tracks the experimental EDC with high fidelity [inset of Fig.~\ref{fig:parameters}(b)]. A comparison of the full bandmap shown in Fig.~\ref{fig:parameters}(b) with the data in Fig.~\ref{fig2}(a) of the main text nevertheless shows a discrepancy of the quasiparticle dispersions, likely because of our non-optimal choice for the bare-band parameters $m_b$ and $S$.

In order to improve our estimate for the bare band, we proceed with a global optimization of all model parameters, by fitting them to a series of 18 EDC curves spanning the momentum range $[-0.4,+0.4]$~\AA{}$^{-1}$. Because the satellite features are strongest at $k_{\parallel}=0$, we constrain the parameters that influence them the most, e.g.\ $\hbar\omega_2$, $\epsilon_{1,2}$, and $\Delta E$ to provide the best fit at $k_{\parallel}=0$. Specifically, the least-square function that we minimize takes all parameters but $\hbar\omega_2$, $\epsilon_{1,2}$, and $\Delta E$ as arguments and involves an internal step, where these parameters are optimized with respect to the EDC at $k_{\parallel}=0$. Since the residual scattering may be momentum dependent, the parameter $\Gamma$ is allowed to vary from one EDC to the next, although we find that the best fit stays very close to $\Gamma=7.7$~meV for all momenta. To represent the effect of a photoemission matrix element, we also allow the overall amplitude to vary with momentum, and we find that it follows a parabolic curve with maximum at $k\approx0$. Finally, we permit a small misalignment correction $\Delta k$, which is added to all experimentally-determined wavevectors, and which the fitting determines as $\Delta k=-0.0039$~\AA$^{-1}$. This fitting procedure involves optimization in a high-dimensional space (51 parameters in total), where the local minimum reached will generally depend on the starting point of the search. Here, we start directly from our set of initial estimated parameters. The minimum reached corresponds to $m_b=2.72m$, $S=3.01$~\AA{}$^2$, $d=3.0$~\AA{}, $N_s=83$, $\hbar\omega_2=164$~meV, $\epsilon_{1,2}=3.7, 7.9$, $g_{\mathrm{A_1}}=0.006$~eV, $g_{\mathrm{LO}}=1.69$~eV, and $\Delta E=58$~meV. The full experimental data set used for fitting and the best fit are displayed in Fig.~\ref{fig:parameters}(d). The resulting simulated ARPES bandmap shown in Fig.~\ref{fig:parameters}(c) is the same as in Fig.~\ref{fig2}(c) of the main text.

\section{\boldmath Comparison of different models for WS$_2$/hBN at $\Gamma$}
\label{app:k=0}

\begin{figure}[tb]
\includegraphics[width=\columnwidth]{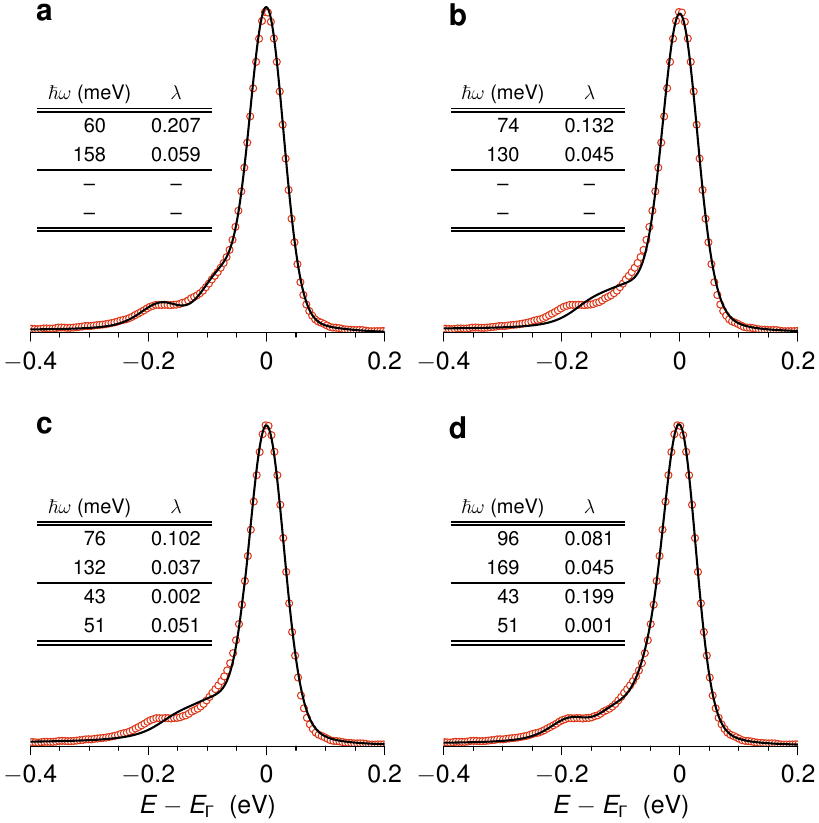}
\caption{\label{fig:k=0}
Best fit of four different models (solid lines) to the WS$_2$/hBN EDC at $\Gamma$ (dots). The first (last) rows in each table show the energies and coupling strengths to hBN (WS$_2$) phonons. \textbf{a}. Only hBN phonons. \textbf{b} Only hBN phonons with fixed $\hbar\omega_1/\hbar\omega_2$. \textbf{c} hBN phonons with fixed $\hbar\omega_1/\hbar\omega_2$ and unoptimized WS$_2$ phonons. \textbf{d} Like \textbf{c}, but with optimized coupling to WS$_2$ phonons.
}
\end{figure}

In this section, we compare how different models fit the WS$_2$/hBN EDC at the $\Gamma$ point. The main observation is that the experimental data are not compatible with a model that only couples to hBN phonons. However, they indicate that the isotropic coupling to the $A_1$ mode of WS$_2$ is weaker than predicted by theory \cite{Sohier2016}. As a first scenario, we ignore the WS$_2$ phonons altogether and we fit the model with two hBN phonons, leaving the phonon energies and coupling strengths free to adjust and the other parameters fixed to their initial values. The fit resembles the experimental data [Fig.~\ref{fig:k=0}(a)]. However, it places the optimal energy of the lowest phonon at $\hbar\omega_1=60$~meV, closer to WS$_2$ optical modes than the hBN ZO mode. This mode comes with a coupling $\lambda_1=0.21$, while the second mode has weaker coupling $\lambda_2=0.06$ and an energy $\hbar\omega_2=158$~meV, still compatible with the hBN LO/TO modes. In a second scenario, we constrain the lowest-mode energy by fixing the ratio $\hbar\omega_1/\hbar\omega_2$ to the value $97/170$ and letting $\hbar\omega_2$ adjust. The result is a bad compromise, where $\hbar\omega_1=74$~meV and $\hbar\omega_2=130$~meV, which yields a poor fit [Fig.~\ref{fig:k=0}(b)]. In a third scenario, we include the coupling to the LO and $A_1$ phonons of WS$_2$ as described in App.~\ref{app:WS2}, using the first-principles values for $g_{\mathrm{LO}}$ and $g_{A_1}$ and letting $\hbar\omega_2$ adjust with $\hbar\omega_1/\hbar\omega_2$ fixed, like previously. This again produces a rather poor fit, where the coupling to the highest hBN mode is suppressed ($\hbar\omega_2=132$~meV with $\lambda_2=0.037$) resulting in a misplaced and too weak satellite feature [Fig.~\ref{fig:k=0}(c)]. The fourth scenario is like the third, except that the coupling strengths to the WS$_2$ phonons are left free to adjust. Strikingly, this version achieves an excellent fit [Fig.~\ref{fig:k=0}(d)] with the expected energies for the hBN phonons, by suppressing the coupling to the $A_1$ mode and increasing the coupling to the LO mode of WS$_2$. The coupling $g_{\mathrm{LO}}$ is increased by a factor $\sim 10$ relative to the first-principles value, such that $\lambda\propto g_{\mathrm{LO}}^2$ is increased by a factor $\sim 100$ to $\approx 0.2$.

The scattering rate associated with the isotropic coupling to the $A_1$ mode has a step at $-51$~meV and stays approximately constant at larger binding energies [see also Fig.~\ref{fig:sigma}(a-2) below]. With the first-principles value of $g_{A_1}$, this scattering rate is already too large in the energy range between $-0.2$ and $-0.1$~eV, leaving no space for the two hBN phonons. Hence the best fit suppresses the coupling to those phonons in Fig.~\ref{fig:k=0}(c). By contrast, the scattering rate associated with the WS$_2$ LO mode has a \emph{peak} at $-43$~meV and drops at larger binding energies [see also Fig.~\ref{fig:sigma}(b-2) below]. Therefore, even if strongly enhanced, it does not compete with the hBN phonons. Our analysis shows that the data at binding energies $\lesssim100$~meV below the quasiparticle peak seems to require forward-scattering electron-phonon interactions to be explained. However, the resolution of the data is not sufficient to fully distinguish WS$_2$ from the hBN ZO phonon in that energy range. The identification of the high energy satellite S2 with coupling to the hBN LO mode, on the other hand, is a robust feature of all fits.

\section{\boldmath Estimated and optimized parameters for WSe$_2$ on hBN}
\label{app:WSe2}

\begin{figure}[tb]
\includegraphics[width=\columnwidth]{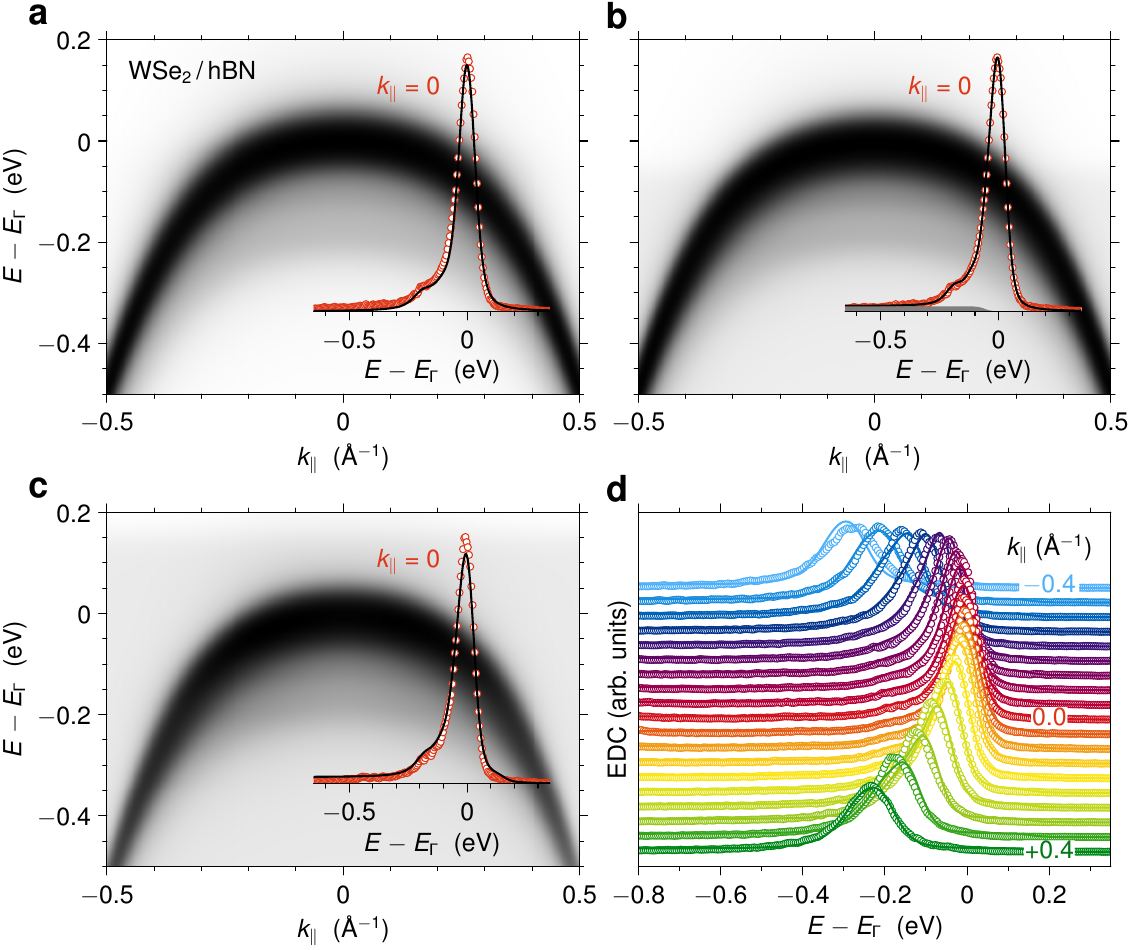}
\caption{\label{fig:WSe2}
\textbf{a} Theoretical ARPES bandmap computed with the initial values of the parameters for WSe$_2$/hBN. The inset shows the experimental EDC at $k_{\parallel}=0$ (red symbols) together with the theoretical curve. \textbf{b} ARPES bandmap computed after some parameters have been adjusted to fit the EDC at $k_{\parallel}=0$. A small background (grey shade) is included. \textbf{c} ARPES bandmap and EDC at $k_{\parallel}=0$ computed after an optimization of all parameters. \textbf{d} Best fit (solid lines) to 19 experimental EDC curves (circles) with a momentum dependent amplitude. The curves are shifted vertically for clarity.
}
\end{figure}

We analyze the data for WSe$_2$/hBN with the same procedure as for WS$_2$/hBN. The initial parameters that are related to the hBN substrate are identical in both cases ($d$, $N_s$, $\hbar\omega_{1,2}$, $\epsilon_{1,2}$). Owing to the near absence of mass enhancement in WS$_2$/hBN, we extract the estimated band mass and quartic correction directly from the WSe$_2$ ARPES bandmap, finding $m_b=4.16m$ and $S=4.22$~\AA{}$^2$. We use the first-principles values given in Ref.~\cite{Sohier2016} for the properties of the two WSe$_2$ phonons: $g_{A_1}=0.08$~eV for the $A_1$ mode at energy $30$~meV and $g_{\vec{q},\mathrm{LO}}(q)=g_{\mathrm{LO}}e^{-qr_0}$ with $g_{\mathrm{LO}}=0.323$~eV and $r_0=48.7$~\AA{} for the LO mode at $30$~meV. As the ECDs for WSe$_2$ are broader than for WS$_2$, we start with a larger $\Gamma=15$~meV, while keeping the same Gaussian broadening $\Delta E=64$~meV. These initial parameters lead to the bandmap and EDC at $k_{\parallel}=0$ shown in Fig.~\ref{fig:WSe2}(a). Figure~\ref{fig:WSe2}(b) shows the result of fitting parameters to the EDC at $k_{\parallel}=0$, which yield an excellent fit with parameters $d=4.5$~\AA{}, $\hbar\omega_2=171$~meV, $\epsilon_{1,2}=3.6, 7.8$, $g_{\mathrm{A_1}}=0.02$~eV, $g_{\mathrm{LO}}=1.7$~eV, $\Gamma=10$~meV, and $\Delta E=59$~meV. The small background visible in the inset has relative amplitude $B_1=0.17~\mathrm{eV}^{-1}$, with $B_2=-59$~meV and $B_3=15$~meV. Figure~\ref{fig:WSe2}(d) shows 19 EDCs fitted together and Fig.~\ref{fig:WSe2}(c) the resulting bandmap and EDC at $k_{\parallel}=0$, with the final parameters $m_b=3.83m$, $S=4.24$~\AA{}$^2$, $d=2.8$~\AA{}, $N_s=93$, $\hbar\omega_2=159$~meV, $\epsilon_{1,2}=4.0, 9.6$, $g_{\mathrm{A_1}}=0.02$~eV, $g_{\mathrm{LO}}=1.84$~eV, $\Gamma\approx11$~meV, and $\Delta E=61$~meV. A small qualitative difference with respect to WS$_2$ is found in the fitted matrix element, which presents a minimum at $k_{\parallel}=0$ and two maxima near $k_{\parallel}=\pm0.15$~\AA{}$^{-1}$.

The optimized parameters for the coupling to the hBN LO mode are very similar to those obtained for WS$_2$ (values for WS$_2$ are indicated in parentheses). The mode is placed at 159~meV (164~meV) with a coupling constant $\alpha_{\mathrm{LO}}=0.16$ (0.17), corresponding to $\lambda_{\mathrm{LO}}=0.038$ (0.048). For the ZO mode at 91~meV (94~meV), we have $\alpha_{\mathrm{ZO}}=0.33$ (0.30), corresponding to $\lambda_{\mathrm{ZO}}=0.091$ (0.091), giving a total couping to the substrate $\lambda_{\mathrm{hBN}}=0.13$ (0.14). Also similarly to WS$_2$, the fit favors the forward-scattering LO mode of WSe$_2$---although less markedly than for WS$_2$---by reducing the coupling to the isotropic $A_1$ mode by a factor 4.0 (14) with respect to the first-principles value, and enhancing the coupling to the LO mode by a factor 5.7 (10), giving $\lambda_{\textrm{A}_1}=0.01$ $(\approx 0)$ and $\lambda_{\textrm{LO}}=0.42$ (0.23).

\section{General properties of the self-energy and spectral function}
\label{sec:properties}

\begin{figure}[tb]
\includegraphics[width=\columnwidth]{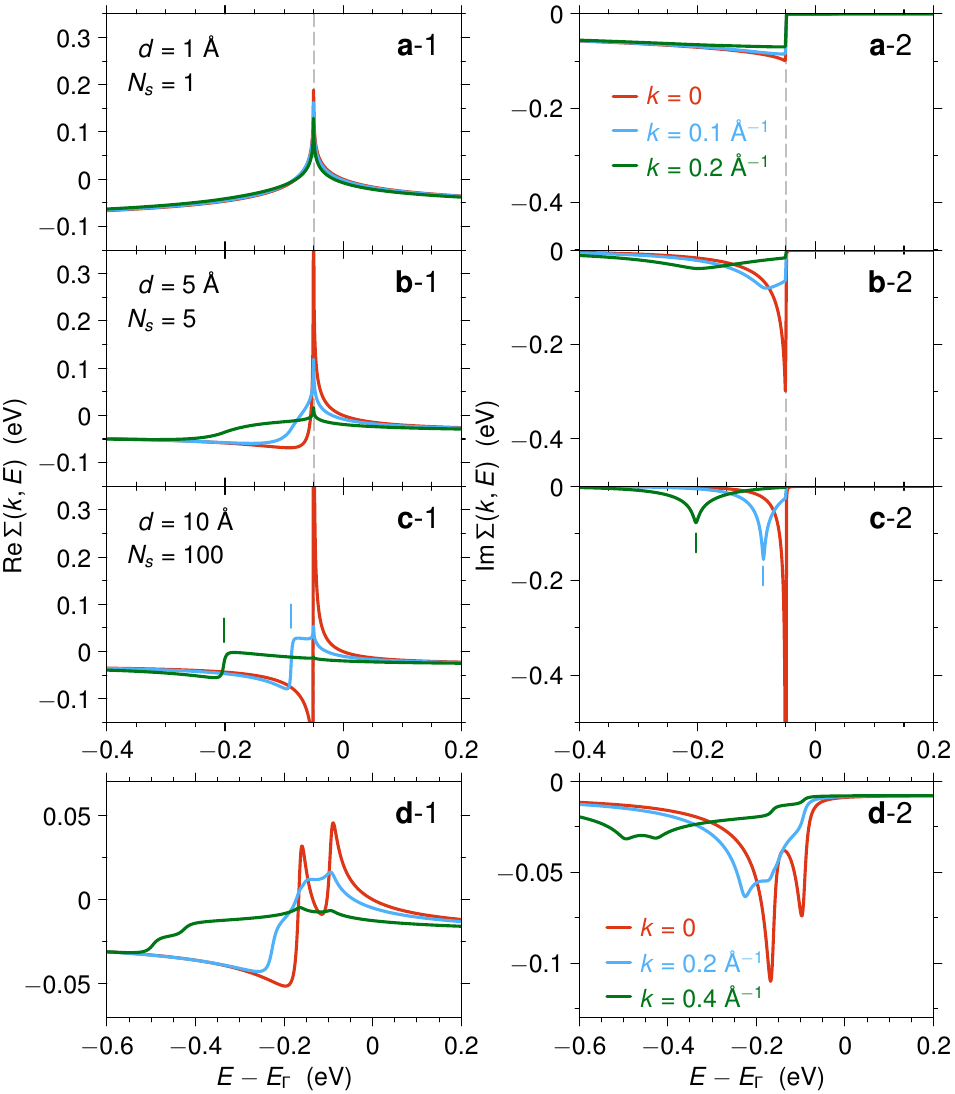}
\caption{\label{fig:sigma}
Real part (left panels) and imaginary part (right panels) of the self-energy vs energy for three wavevectors $k$. Panels \textbf{a}, \textbf{b}, and \textbf{c} display the self-energy for a parabolic band of mass $m_b=m$ and a single phonon of energy $\hbar\omega_0=0.05$~eV, a coupling strength $\lambda=0.5$, and \textbf{a} $d=1$~\AA{}, $N_s=1$, \textbf{b} $d=5$~\AA{}, $N_s=5$, and \textbf{c} $d=10$~\AA{}, $N_s=100$. Panels \textbf{d} display the self-energy due to the hBN modes with the parameters optimized for WS$_2$/hBN, which corresponds to $\lambda=0.14$. The dashed gray lines indicate the energy of the phonon and the vertical bars in \textbf{c} mark the energy $\hbar^2k^2/(2m)$ measured from the phonon energy.
}
\end{figure}

Depending on the values of $d$ and $N_s$, the model defined by Eq{}s.~(\ref{eq:Sigma}) and (\ref{eq:gbar}) interpolates between the limiting case of a purely local ($k$-independent) self-energy with Migdal-Eliashberg type of behavior, which produces dispersion kinks in the spectral function, and that of a polaronic self-energy, which produces satellite peaks. We illustrate this evolution in Fig.~\ref{fig:sigma} for a free-hole band and a single phonon with $\hbar\omega_0=50$~meV, using a small broadening $\delta=0.1$~meV.

\begin{figure}[tb]
\includegraphics[width=\columnwidth]{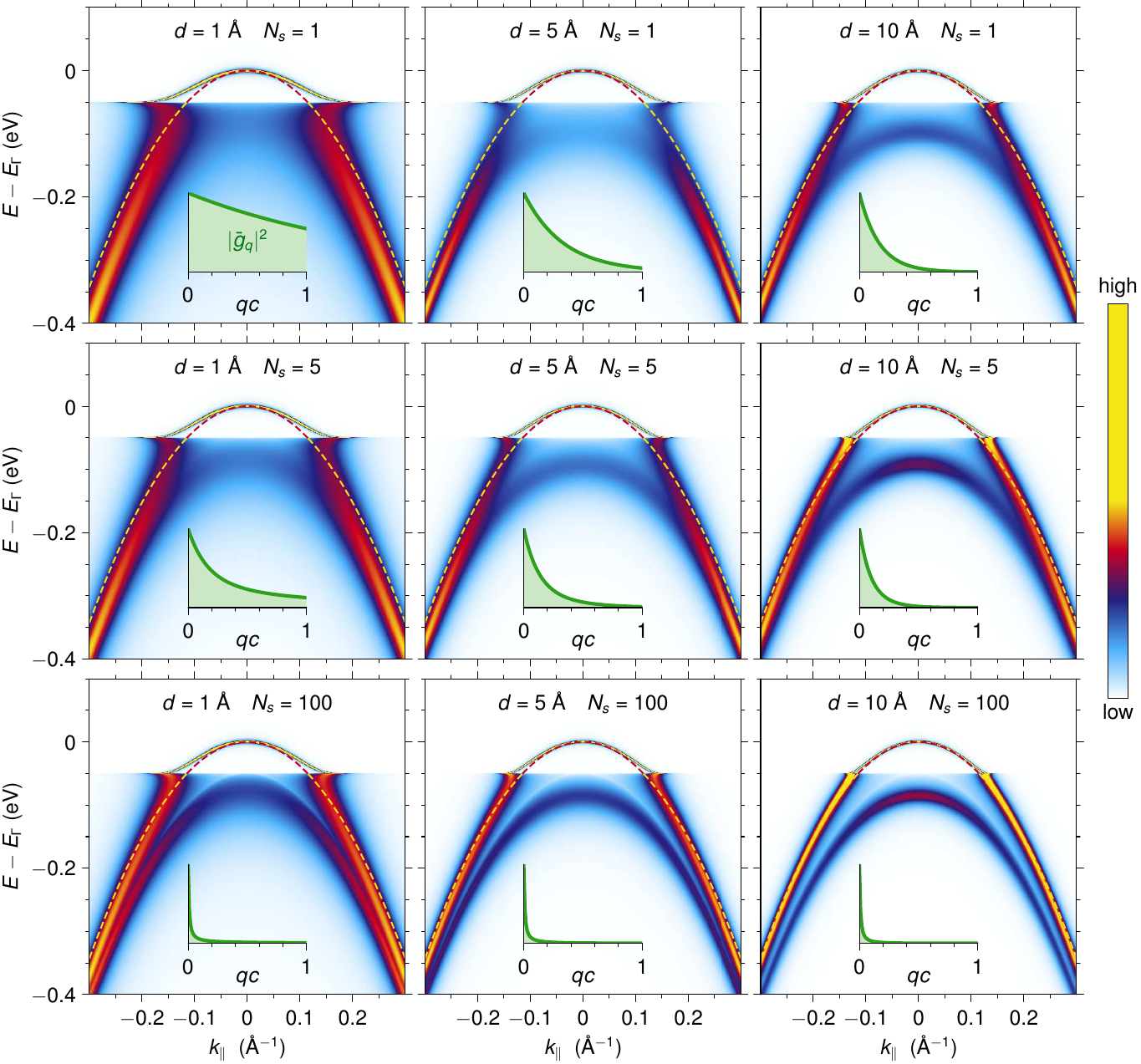}
\caption{\label{fig:A}
Spectral function $A(k,E)$ for a parabolic band of mass $m_b=m$ and a single phonon of energy $\hbar\omega_0=0.05$~eV, a coupling strength $\lambda=0.5$, and various values of $d$ and $N_s$. The small broadening of $A(k,E-E_{\Gamma}>-\hbar\omega_0)$ reflects the value $\delta=0.1$~meV. The dashed line in each panel shows the bare band and the inset displays the coupling function $\bar{g}^2_{\vec{q}}$. 
}
\end{figure}

When $d$ and $N_s$ are both small, the coupling function $\bar{g}_{\vec{q}}^2$ approaches a $\vec{q}$-independent function (see also Fig.~\ref{fig:A}). In the limit of a constant $\bar{g}_{\vec{q}}^2$, a hole at wavevector $\vec{k}$ with energy exceeding $\hbar\omega_0$ can excite phonons of arbitrary wavevectors $\vec{q}$. The resulting scattering rate [Fig.~\ref{fig:sigma}(a-2)] presents a momentum-independent step at $E=-\hbar\omega_0$, with the behavior above the step following the density of states of the band---which is constant for a 2D parabolic band. The corresponding real part has a logarithmic singularity at the energy of the step [Fig.~\ref{fig:sigma}(a-1)].

At the other end, when $d$ and $N_s$ are both large, $\bar{g}_{\vec{q}}^2$ is strongly peaked at $q=0$. In the extreme limit where $\bar{g}_{\vec{q}}^2$ would be proportional to a Dirac delta function at $q=0$, the scattering rate would be a delta function at $E=\xi_{\vec{k}}-\hbar\omega_0$ [see Eq.~(\ref{eq:Sigma}) and the vertical bars in Fig.~\ref{fig:sigma}(c-2)], meaning that the only permitted relaxation process at wavevector $\vec{k}$ is a momentum-conserving phonon emission. Such a transition is impossible from a bare band, but becomes possible in the presence of a sideband at energy $\xi_{\vec{k}}-\hbar\omega_0$. The emergence of this sideband results from the singular energy renormalization $\sim\bar{g}_{\vec{0}}^2/(E-\xi_{\vec{k}}+\hbar\omega_0)$---see Fig.~\ref{fig:sigma}(c-1)---which ensures that the quasiparticle equation $E_{\vec{k}}-\xi_{\vec{k}}-\mathrm{Re}\,\Sigma(\vec{k},E_{\vec{k}})$ admits two solutions, which are $E_{\vec{k}}=\xi_{\vec{k}}$ and $E_{\vec{k}}=\xi_{\vec{k}}-\hbar\omega_0$ in the weak-coupling limit.

Fig.~\ref{fig:sigma}(b) illustrates the behavior for intermediate values of $d$ and $N_s$, while Fig.~\ref{fig:sigma}(d) shows the two-phonon self-energy with parameters optimized for WS$_2$/hBN.

The spectral functions $A(k,E)$ computed for a free-hole band, a single phonon, and various values of $d$ and $N_s$, are displayed in Fig.~\ref{fig:A}. If $\delta=0^+$, $A(k,E)$ is infinitely sharp at $E-E_{\Gamma}>-\hbar\omega_0$, where $\Sigma(k,E)=-i\delta$. With our choice, $\delta=0.1$~meV, the low-energy quasiparticle appears as a narrow band with saturated color. When $d$ and $N_s$ are both small, we observe the typical dispersion kink at the junction of the low-energy and high-energy quasiparticle branches, with only a faint trace of a sideband. Keeping $d$ small and increasing $N_s$, the kink weakens and the sideband sharpens, such that both of them are present in the bulk limit $N_s\to\infty$. A similar evolution occurs upon increasing $d$ at small $N_s$, although this regime is slightly artificial, since it requires boosting the Coulomb interaction to maintain a constant coupling strength as the substrate goes away. When $d$ and $N_s$ are both large, the kink is hardly noticeable and all the displaced spectral weight goes to the sideband. Note that the energy shift of the sideband relative to the bare band is larger than $\hbar\omega_0$ in Fig.~\ref{fig:A}. This is a consequence of pushing the leading-order self-energy Eq.~(\ref{eq:Sigma}) to a relatively large coupling strength $\lambda=0.5$.

Fig.~\ref{fig:A} reveals that, although all panels share a common coupling strength $\lambda=0.5$, they present different mass renormalizations. In the theories with momentum-independent self-energy, the mass renormalization is $m^*/m=1+\lambda$. This renormalization is indeed visible for $d=1$~\AA{} and $N_s=1$, a case where the self-energy is almost independent of $k$ (see Fig.~\ref{fig:sigma}). For $d=10$~\AA{} and $N_s=100$, however, no noticeable mass renormalization is visible. We show in the next section that the absence of mass enhancement is indeed a consequence of the momentum dependence of the self-energy.

\section{Suppression of mass enhancement in the forward-scattering regime}
\label{app:m*/m}

A momentum-independent self-energy $\Sigma(E)$ yields a mass enhancement $m^*/m=1+\lambda$ \cite{Mahan-2000, Berthod-2018}, where the renormalization factor $\lambda=-d\mathrm{Re}\,\Sigma(E)/dE$ is evaluated at the energy where the mass is defined, i.e., the top of the valence band at the $\Gamma$ point in the present case. The self-energy in Eq.~(\ref{eq:Sigma}) depends on momentum, which changes the mass renormalization. Here, we show that for the model Eq.~(\ref{eq:Sigma}), the momentum dependence of the self-energy nearly cancels the mass renormalization stemming from the energy dependence. We specialize to the case of an isotropic band, where the self-energy is isotropic as well.

The effective mass measures the curvature of the quasiparticle dispersion $E_k$ according to $\hbar^2/m^*=d^2E_k/dk^2$, where the derivative is evaluated at the $\Gamma$ point in the present case. As the imaginary part of the self-energy vanishes at and near the $\Gamma$ point, the quasiparticle energy $E_k$ is the solution of the implicit equation $E_k-\xi_k-\mathrm{Re}\,\Sigma(k,E_k)=0$. Because this equation is satisfied at all $k$ points in the neighborhood of $\Gamma$, we can differentiate it twice with respect to $k$ and deduce expressions for the quasiparticle velocity and for the curvature,
\begin{align*}
	\frac{dE_k}{dk}&=\frac{1}{1-\frac{\partial\Sigma_1}{\partial E}}\left(\frac{d\xi_k}{dk}
	+\frac{\partial\Sigma_1}{\partial k}\right)\\
	\frac{d^2E_k}{dk^2}&=\frac{1}{1-\frac{\partial\Sigma_1}{\partial E}}\left[\frac{d^2\xi_k}{dk^2}+
	\frac{\partial^2\Sigma_1}{\partial k^2}+2\frac{\partial^2\Sigma_1}{\partial k\partial E}
	\frac{dE_k}{dk}\phantom{\left(\frac{1_k}{1}\right)^2\kern-3em}\right.\\
	&\quad\left.+\frac{\partial^2\Sigma_1}{\partial E^2}\left(\frac{dE_k}{dk}\right)^2\right],
\end{align*}
where $\Sigma_1\equiv\mathrm{Re}\,\Sigma(k,E)$ and the right-hand side is evaluated at $E=E_k$. At the $\Gamma$ point of the valence band, $dE_k/dk=0$ by symmetry and we deduce the mass enhancement
\begin{subequations}\begin{align}
	\label{eq:mstar}
	\frac{m^*}{m}&=\frac{1+\lambda}{1+\rho}\\
	\lambda&=-\left.\frac{\partial\Sigma_1}{\partial E}\right|_{k=0, E=E_{\Gamma}}\\
	\rho&=\left.\frac{\partial^2\Sigma_1/\partial k^2}{d^2\xi_k/dk^2}\right|_{k=0, E=E_{\Gamma}}.
\end{align}\end{subequations}
We proceed to estimate $\lambda$ and $\rho$ in the case of a parabolic hole band $\xi_k=E_{\Gamma}-\frac{\hbar^2k^2}{2m_b}$. In this case, we find that the energy renormalization factor is
\begin{multline}\label{eq:lambda}
	\lambda=\frac{4\alpha}{\pi}\left(\frac{\hbar\omega_0}{E_c}\right)^{\frac{3}{2}}
	\int_0^{\infty}dx\,\frac{1}{(\hbar\omega_0/E_c+x^2)^2}\\
	\times xe^{-2xd/c}\frac{1-e^{-2xN_s}}{1-e^{-2x}},
\end{multline}
where we have defined the energy $E_c=\hbar^2/(2m_bc^2)$.
%
%
For the momentum renormalization, after performing the angle integration in Eq.~(\ref{eq:Sigma}) and taking the second derivative with respect to $k$, we arrive at
\begin{multline}\label{eq:rho}
	\rho=\frac{4\alpha}{\pi}\left(\frac{\hbar\omega_0}{E_c}\right)^{\frac{3}{2}}
	\int_0^{\infty}dx\,\frac{\hbar\omega_0/E_c-x^2}{(\hbar\omega_0/E_c+x^2)^3}\\
	\times xe^{-2xd/c}\frac{1-e^{-2xN_s}}{1-e^{-2x}}.
\end{multline}
For $N_s\gg1$, the factor $xe^{-2xd/c}/(1-e^{-2x})$ confines $x$ to the range $x\lesssim 2c/d$ for both integrals in Eq{}s.~(\ref{eq:lambda}) and (\ref{eq:rho}). We therefore see that if $\hbar\omega_0/E_c\gg(2c/d)^2$, in other words if $\hbar\omega_0\gg2\hbar^2/(m_bd^2)$, both integrals approach $(\hbar\omega_0/E_c)^{-2}\int_0^{\infty}dx\,\frac{xe^{-2xd/c}}{1-e^{-2x}}$ and thus $\lambda=\rho$ and $m^*=m$ in this limit. With the parameters appropriate for WS$_2$/hBN, however, $\hbar\omega_0$ and $2\hbar^2/(m_bd^2)$ are similar. A numerical evaluation with $c=3.33$~\AA{} and $d=5$~\AA{} shows that $\rho/\lambda$ remains nevertheless close to unity. With $\lambda$ and $\rho$ being nearly equal, the self-energy Eq.~(\ref{eq:Sigma}) brings almost no mass renormalization.

\section{Comparison with the Fr\"{o}hlich polaron}
\label{app:Z}

The polaron theory of Fr\"{o}hlich describes an electron of mass $m$ in interaction via the Coulomb force with a polarizable dielectric medium characterized by a single optical mode of frequency $\omega_0$ \cite{Frohlich-1950}. All properties of the Fr\"{o}hlich polaron can be studied as universal functions of a dimensionless coupling constant $\alpha$, while the two other parameters of the theory---electron mass and phonon frequency---are lumped into the length and energy units \cite{Mishchenko2000}. In particular, it is known that the residue $Z$ and the mass enhancement $m^*/m$ of the polaron behave as $Z=1-\alpha/2$ and $m^*/m=(1-\alpha/6)^{-1}$ at weak coupling $\alpha\lesssim1$. Because these quantities are dimensionless, they do not depend on the choice of units, hence the universality. Our model lacks this universality due to additional parameters, in particular the distance $d$ between the substrate and the 2DEG and the number $N_s$ of layers in the substrate. Nonetheless, we expect to find properties similar to those of the Fr\"{o}hlich polaron at weak coupling and for $N_s\gg1$, where our coupling function $\sim 1/q$ resembles the strong forward-scattering condition of the polaron model.

\begin{figure}[tb]
\includegraphics[width=\columnwidth]{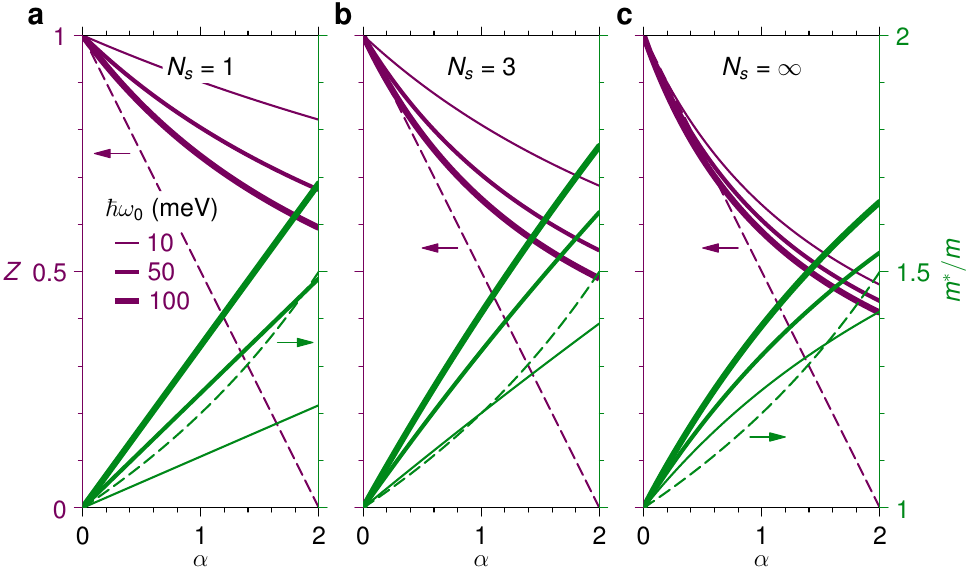}
\caption{\label{fig:Z}
Quasiparticle residue (left axes) and mass enhancement (right axes) versus $\alpha$. The dashed lines show the weak-coupling behavior for the Fr\"{o}hlich polaron. The solid lines show the results obtained with Eq{}s.~(\ref{eq:mstar}), (\ref{eq:lambda}) and (\ref{eq:rho}) with three different phonon frequencies for \textbf{a} $N_s=1$, \textbf{b} $N_s=3$, and \textbf{c} $N_s=\infty$. The other parameters are $m_b=m$ and $d=0$.
}
\end{figure}

The three panels of Fig.~\ref{fig:Z} show the weak-coupling $Z$ and $m^*/m$ of the polaron theory versus $\alpha$ as dashed lines referred to the left and right axes, respectively. The other curves correspond to the residue and mass enhancement calculated with our model for three different phonon energies and increasing values of $N_s$. We set $d=0$, which puts the 2DEG at the surface of a semi-infinite crystal in the limit $N_s\gg1$. Our quasiparticle residue is generally larger than for the polaron and displays a dependence on $\hbar\omega_0$ that weakens as $N_s$ increases.

In our model, the behavior of $Z=(1+\lambda)^{-1}$ may be analyzed as follows. Using Eq{}.~(\ref{eq:lambda}) and defining $\varpi=\sqrt{\hbar\omega_0/E_c}$, we rewrite $\lambda=2\alpha\int_0^{\infty}dx\,S(x)G(x)$, where $S(x)=(2/\pi)\varpi^3/(\varpi^2+x^2)^2$ and $G(x)=x(1-e^{-2xN_s})/(1-e^{-2x})$ for $d=0$. In these expressions, $\varpi$ represents the phonon frequency and $x$ the exchanged momentum $q$. The function $S(x)$ describes the phase space available for scattering, which is narrowly concentrated around $x=0$ for small $\varpi$ and spreads more for larger $\varpi$, irrespective of the value of $N_s$. In the limit $\varpi\to0$, $S(x)$ approaches the Dirac delta function $\delta(x)$. From this, we deduce that $\lambda=\alpha G(0)$ in the limit of small phonon frequency. The function $G(x)$ represents the interaction kernel, which is qualitatively different for $N_s=1$, where it becomes $x$ for $x\to0$, and for $N_s=\infty$, where it becomes $(1-x)/2$. This difference illustrates the effect of piling up the Coulomb interaction from a semi-infinite number of layers, rather than a single layer (see App.~\ref{app:self}). If follows that, for $\varpi\to0$, we have $\lambda=0$ if $N_s=1$ and $\lambda=\alpha/2$ if $N_s=\infty$. Hence $Z$ approaches $1$ in the former case, as illustrated by the thin purple curve in Fig.~\ref{fig:Z}(a), while it approaches $1-\alpha/2$ in the latter, like for the Fr\"{o}hlich polaron [thin purple curve in Fig.~\ref{fig:Z}(c)].

The analysis of the mass enhancement is done similarly using $m^*/m=(1+\lambda)/(1+\rho)$, with the function $S(x)$ entering the definition of $\rho$ now approaching $\delta(x)/2$. One thus finds $\rho=0$ and $\rho=\alpha/4$ for $N_s=1$ and $N_s=\infty$, respectively, in the limit $\hbar\omega_0\to0$. Hence the enhancement disappears for $N_s=1$ and it approaches $(1+\alpha/2)/(1+\alpha/4)$ at weak coupling for $N_s=\infty$.

The correspondence of our model with the Fr\"{o}hlich polaron regarding $Z$ was imposed by an appropriate definition of $\alpha$ in Eq.~(\ref{eq:gbar1}). Now we compare, for a thick substrate and for $q\to0$, our interaction vertex $\bar{g}_{\vec{q}}^2$ with that of the polaron theory. For $N_s=\infty$, we have
\begin{equation}
	\bar{g}_{\vec{q}}^2=\alpha\sqrt{8}\sqrt{\frac{\hbar^5\omega_0^3}{m_b}}\frac{1}{q}\quad\mathrm{for}\quad q\to0.
\end{equation}
In the Fr\"{o}hlich model, the electron-phonon matrix element is written as $|V(\vec{p})|^2=2\sqrt{2}\alpha\pi\frac{1}{p^2}m_b\hbar\omega_0^3$ \cite{Mishchenko2000}. We have included the factor $m_b\hbar\omega_0^3$, which ensures that $|V(\vec{p})|^2$ is an energy, considering that Ref.~\cite{Mishchenko2000} sets $m_b=\hbar=\omega_0=1$. In order to compare with our expression for $\bar{g}_{\vec{q}}^2$, we must average $|V(\vec{p})|^2$ over the $p_z$ component of the three-dimensional vector $\vec{p}$, which amounts to replacing $1/p^2$ by $1/2p_{\parallel}=1/2q$. We must also multiply $|V(\vec{p})|^2$ by a unit normalization volume, which is $[\hbar/(m_b\omega_0)]^{3/2}$ in the system of units used in Ref.~\cite{Mishchenko2000}. Performing these transformations, we find
\begin{equation}
	\left(\frac{\hbar}{m_b\omega_0}\right)^{\frac{3}{2}}\int_{-\infty}^{\infty}\frac{dq_z}{2\pi}|V(\vec{q})|^2
	=\alpha\pi\sqrt{2}\sqrt{\frac{\hbar^5\omega_0^3}{m_b}}\frac{1}{q}.
\end{equation}
Hence our interaction $V(q\to0)$ is slightly weaker than the Fr\"{o}hlich-polaron interaction, by a factor $\sqrt{\pi/2}$, which must be ascribed to the different geometries.

\section{Causal Fermi-liquid self-energy for WS$_2$ on graphite}
\label{app:FL}

In a Fermi liquid, the single-particle excitations relax by generating particle-hole pairs across the Fermi surface. At low excitation energies $E$ and temperatures $T$, the corresponding scattering rate behaves as $\mathrm{Im}\,\Sigma_{\mathrm{FL}}(E,T)\propto E^2+(\pi k_{\mathrm{B}}T)^2$. Our goal is to build a causal (i.e., Kramers--Kronig consistent) phenomenological self-energy model that exhibits a Fermi-liquid behavior at low energy and temperature. For this, we must cut the unbounded $E^2$ increase of the scattering rate, which violates causality at high energy. We therefore introduce a soft cutoff at the characteristic energy $W$, above which the scattering rate vanishes as $1/E^2$. We then perform the Kramers--Kronig integral to deduce the real part. The resulting model has two parameters, the cutoff $W$ and an overall amplitude that can be parametrized by the dimensionless coupling $\lambda_{\mathrm{FL}}=-d\mathrm{Re}\,\Sigma_{\mathrm{FL}}(E,T)/dE|_{E=0}$. We thus get \cite{Berthod-2018}
\begin{multline}
	\Sigma_{\mathrm{FL}}(E,T)=
	\frac{\lambda_{\mathrm{FL}} W}{1-\tilde{T}^2}\left[-\tilde{E}
	\frac{1-\tilde{T}^2-(1+\tilde{T}^2)\tilde{E}^2}{1+\tilde{E}^4}\right.\\\left.
	-i\sqrt{2}\frac{\tilde{E}^2+\tilde{T}^2}{1+\tilde{E}^4}\right]
\end{multline}
with $\tilde{E}=E/W$ and $\tilde{T}=\pi k_{\mathrm{B}}T/W$. In the WS$_2$/graphite system, the single-particle excitations at the top of the valence band of WS$_2$ relax by emitting particle-hole pairs in graphite. If $E_0$ is the energy of the (unrenormalized) valence-band maximum and $E_{\mathrm{F}}$ the Fermi energy in graphite, the WS$_2$ spectral function at the $\Gamma$ point may be modeled as
\begin{equation}
	A(\Gamma,E)=-\frac{1}{\pi}\mathrm{Im}\,\frac{1}
	{E-E_0-\Sigma_{\mathrm{FL}}(E-E_{\mathrm{F}},T)}.
\end{equation}
Figure~\ref{fig2}(h) of the main text shows a fit of this expression to the EDC of WS$_2$/graphite at the $\Gamma$ point, after a further convolution with a Gaussian of full width at half maximum $\Delta E=35.5$~meV. The graphite Fermi energy was set to $E_{\mathrm{F}}=E_{\Gamma}+1.35$~eV, where $E_{\Gamma}=E_0+66$~meV is the renormalized valence-band maximum. The resulting parameters are $\lambda_{\mathrm{FL}}=0.055$, and $W=4.22$~eV.

\bibliography{MLTMD_hBN, supplementary}

\end{document}